\documentclass[11pt,letterpaper]{article}

\usepackage{threeparttable}
\usepackage{booktabs}
\usepackage{multirow}
\usepackage{pdflscape}
\usepackage{comment}
\usepackage{pdfpages}
\usepackage{fancyhdr}
\usepackage{listings}
\usepackage[table,xcdraw,dvipsnames]{xcolor}
\usepackage{color,colortbl}
\usepackage{titlesec}
\usepackage{times}
\usepackage{relsize}
\usepackage{acro}
\usepackage[numbers]{natbib}
\usepackage{geometry}

\AtBeginDocument{
  \definecolor{pdfurlcolor}{HTML}{037106}
  \definecolor{pdfcitecolor}{rgb}{0,0.1,0}
  \definecolor{pdflinkcolor}{HTML}{037106}
  \definecolor{light}{gray}{.85}
  \definecolor{vlight}{gray}{.95}
}
\usepackage[colorlinks=true,citecolor=pdfcitecolor,urlcolor=pdfurlcolor,linkcolor=pdflinkcolor,pdfborder={0 0 0}]{hyperref}

\titlespacing{\section}{0pc}{1pc}{0.5pc}
\titlespacing{\subsection}{0pc}{1pc}{0.5pc}
\titlespacing{\subsubsection}{0pc}{0.5pc}{0.5pc}
\titleformat*{\section}{\Large\bfseries}
\titleformat*{\subsection}{\large\bfseries}
\titleformat*{\subsubsection}{\normalsize\bfseries}

\lstdefinestyle{wcsStyle}{
  columns=fullflexible,
  tabsize=2,
  showspaces=false,
  showstringspaces=false,
  aboveskip=0em,
  belowskip=0em,
}

\DeclareAcronym{ai}{
  short = AI ,
  long = Artificial Intelligence
}

\DeclareAcronym{apis}{
  short = APIs ,
  long = Application Programming Interfaces
}

\DeclareAcronym{asc}{
    short = AmSC ,
    long = American Science Cloud
}

\DeclareAcronym{cwl}{
  short = CWL ,
  long = Common Workflow Language
}

\DeclareAcronym{doe}{
  short = DOE ,
  long = Department of Energy
}

\DeclareAcronym{dois}{
  short = DOIs ,
  long = Digital Object Identifiers
}

\DeclareAcronym{elixir}{
    short = ELIXIR ,
    long = European life-sciences infrastructure for biological information
}

\DeclareAcronym{eosc}{
  short = EOSC,
  long = European Open Science Cloud
}

\DeclareAcronym{eosdis}{
  short = EOSDIS ,
  long = Earth Observing System Data and Information System
}

\DeclareAcronym{fair}{
  short = FAIR ,
  long = {``Findable, Accessible, Interoperable, Reusable''}
}

\DeclareAcronym{fasst}{
  short = FASST ,
  long = Framework for Accelerating Science and Scientific Training
}

\DeclareAcronym{hpc}{
  short = HPC ,
  long = High Performance Computing
}

\DeclareAcronym{iri}{
  short = IRI ,
  long = Integrated Research Infrastructure
}

\DeclareAcronym{llm}{
    short = LLM ,
    long = Large Language Model
}

\DeclareAcronym{ml}{
  short = ML ,
  long = Machine Learning
}

\DeclareAcronym{nairr}{
  short = NAIRR ,
  long = National Artificial Intelligence Research Resource
}

\DeclareAcronym{nrc}{
  short = NRC ,
  long = Nuclear Regulatory Commission
}

\DeclareAcronym{olcf}{
  short = OLCF ,
  long = Oak Ridge Leadership Computing Facility
}

\DeclareAcronym{ornl}{
  short = ORNL ,
  long = Oak Ridge National Laboratory
}

\DeclareAcronym{pids}{
  short = PIDs ,
  long = Persistent Identifiers
}

\DeclareAcronym{rbac}{
  short = RBAC ,
  long = Role-Based Access Control
}

\DeclareAcronym{s3m}{
    short = S3M ,
    long = Secure Scientific Service Mesh
}

\DeclareAcronym{tnr}{
  short = TNR ,
  long = Times New Roman
}

\DeclareAcronym{tm}{
  short = TM ,
  long = Technical Memo
}

\DeclareAcronym{us}{
  short = US ,
  long = United States
}

\DeclareAcronym{wdl}{
    short = WDL ,
    long = Workflow Description Language
}

\DeclareAcronym{wms}{
  short = WMS ,
  long = workflow management system
}

\begin{document}

\includepdf[pages=-]{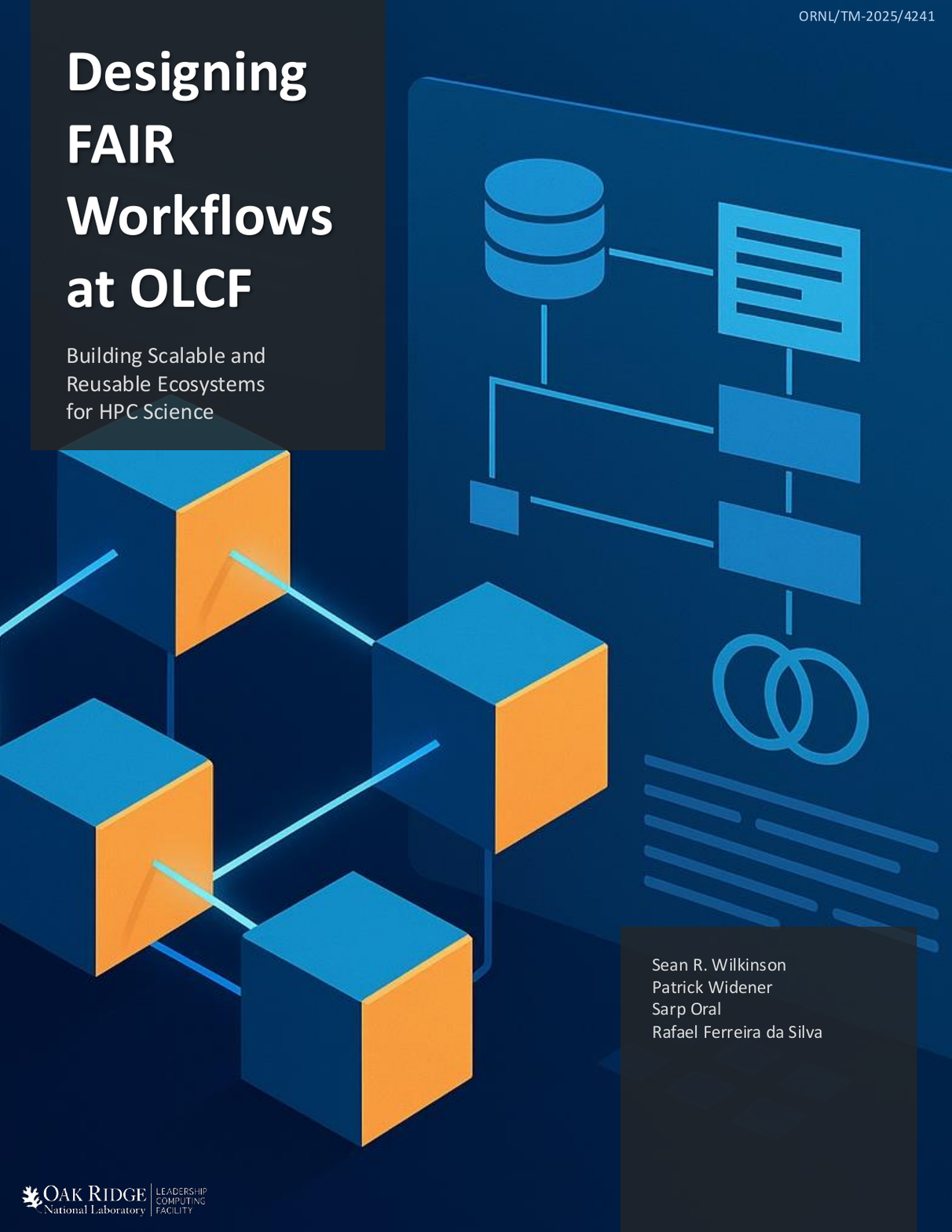}

\pagestyle{fancy}
\fancyhf{}
\rhead{
  \includegraphics[height=11pt]{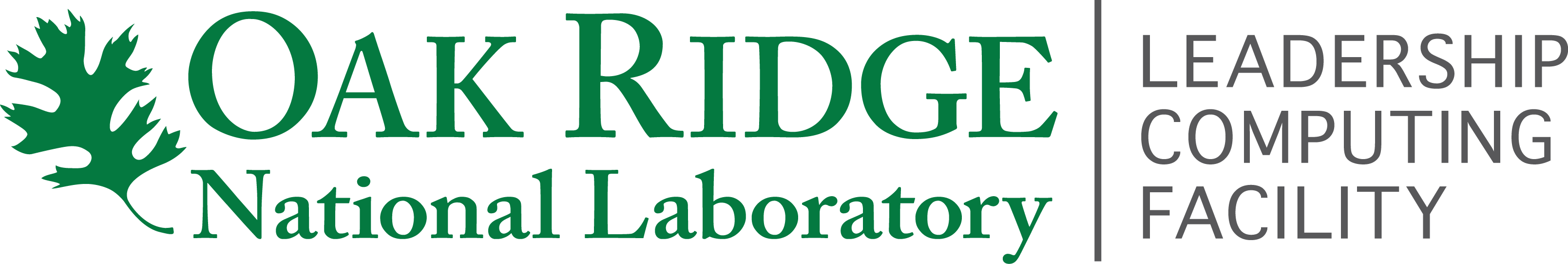}
}
\lhead{\color{gray}\smaller \uppercase{Designing FAIR Workflows at OLCF}}
\rfoot{\thepage}


\begin{table}[!ht]
\centering
\smaller
\begin{tabular}{p{16cm}}
    \textbf{Disclaimer.}
    This research used resources of the Oak Ridge Leadership Computing Facility at ORNL, which is supported by the Office of Science of the U.S. Department of Energy under Contract No. DE-AC05-00OR22725. ORNL is managed by UT-Battelle LLC on behalf of the U. S. Department of Energy.
    \\ 
    \vspace{-0.25em}
    This report was prepared as an account of work sponsored by agencies of the United States Government. Neither the United States Government nor any agency thereof, nor any of their employees, makes any warranty, express or implied, or assumes any legal liability or responsibility for the accuracy, completeness, or usefulness of any information, apparatus, product, or process disclosed, or represents that its use would not infringe privately owned rights. Reference herein to any specific commercial product, process, or service by trade name, trademark, manufacturer, or otherwise, does not necessarily constitute or imply its endorsement, recommendation, or favoring by the United States Government or any agency thereof. The views and opinions of authors expressed herein do not necessarily state or reflect those of the United States Government or any agency thereof.
    \\
    \vspace{-0.25em}
    \textbf{License.}
    This report is made available under a Creative Commons Attribution 4.0 International Public license ({\small \url{https://creativecommons.org/licenses/by/4.0}}).
\end{tabular}
\end{table}

\vspace{-0.5em}
\begin{table}[!ht]
\centering
\smaller
\begin{tabular}{p{16cm}}
    \textbf{\small Preferred citation} 
    \\
    S. R. Wilkinson, P. Widener, S. Oral, R. Ferreira da Silva,
    ``\emph{Designing FAIR Workflows at OLCF: Building Scalable and Reusable Ecosystems for HPC Science}", Technical Report, ORNL/TM-2025/4241, November 2025, DOI: 10.5281/zenodo.17290392.
    \\
    \rowcolor[HTML]{F7F7F7}
    \lstset{basicstyle=\scriptsize,style=wcsStyle}
    \begin{lstlisting}
@techreport{fair-olcf-2025,
  author      = {Wilkinson, Sean R. and Widener, Patrick and Oral, Sarp and Ferreira da Silva, Rafael},
  title       = {{Designing FAIR Workflows at OLCF: Building Scalable and Reusable Ecosystems for HPC Science}},
  year        = {2025},
  publisher   = {Zenodo},
  number      = {ORNL/TM-2025/4241},
  doi         = {10.5281/zenodo.17290392},
  url         = {https://doi.org/10.5281/zenodo.17290392},
  institution = {Oak Ridge National Laboratory}
}
    \end{lstlisting}
    \\
\end{tabular}
\end{table}

\newpage

\tableofcontents
\newpage

\definecolor{note}{RGB}{48,96,192}
\renewcommand{\marginpar}[1]{{\small\em\color{note}[#1]}}


\acresetall 
\section{INTRODUCTION AND MOTIVATION}

\Ac{hpc} centers, such as the \ac{olcf}, provide advanced infrastructure that enables scientific research at extreme scale. These centers operate with unique hardware configurations, specialized software environments, and elevated security requirements that differ substantially from what most users encounter on their local systems. As a result, users often develop customized digital artifacts that are tightly coupled to the specific configuration of a given HPC center. Although necessary, this practice can lead to significant duplication of effort as multiple users independently create similar solutions to common problems.

The FAIR Principles, which stand for Findable, Accessible, Interoperable, and Reusable, offer a framework to address these challenges (Section~\ref{sec:fair}). Initially designed to improve data stewardship, the FAIR approach has since been extended to encompass software, workflows, models, and infrastructure. By encouraging the use of rich metadata and community standards, FAIR practices aim to make digital artifacts easier to share and reuse, both within and across scientific domains.

Many FAIR initiatives have emerged within individual research communities, often aligned by discipline, such as bioinformatics or earth sciences. These communities have made progress in adopting FAIR practices, but their domain-specific nature can lead to silos that limit broader collaboration. To overcome this, we propose that HPC centers play a more active role in fostering FAIR ecosystems that support research across multiple disciplines. This requires designing infrastructure that enables researchers to discover, share, and reuse computational components more effectively.

In this report, we build on the architecture of the \ac{eosc} EOSC-Life FAIR Workflows Collaboratory~\cite{goble_implementing_2021} to propose a model that is tailored to the needs of HPC environments. Rather than focusing on entire workflows, which are often difficult to reuse due to rapid changes in HPC infrastructure, we emphasize the importance of making individual workflow components FAIR. This component-based approach better supports the diverse and evolving needs of HPC users while maximizing the long-term value of their work.

The sections that follow present a conceptual architecture for implementing a FAIR ecosystem at \ac{olcf}, describe existing capabilities and gaps, and outline a strategy to encourage adoption among the user community.

\section{BACKGROUND}
\label{sec:background}
%
To design effective FAIR ecosystems for HPC environments, it is important to understand the foundational principles and existing efforts that inform this work. This section introduces the FAIR Principles, explores how various research communities have implemented them, and outlines specific challenges that arise in the context of HPC centers like OLCF. We also review related initiatives within the US Department of Energy and examine lessons learned from the EOSC-Life FAIR Workflows Collaboratory, which serves as a model for our proposed approach.

\subsection{THE FAIR PRINCIPLES}
\label{sec:fair}

The \ac{fair} Principles for making digital objects Findable, Accessible, Interoperable, and Reusable were first established to improve the management and stewardship of data~\cite{wilkinson2016}. There have been a number of other efforts in this space, some of which even sound similar or related to the word ``fair'', such as the CARE Principles~\cite{carroll2020}, the Fair-code software model\footnote{\url{https://faircode.io/}}, the TRUST Principles~\cite{lin2020}, and several distinct efforts which use the name ``SHARE Principles''~\cite{share2018, jordan2024}. What makes \ac{fair} unique and distinguishes it from other peer initiatives is that, rather than making data easy for just humans to use, it places specific emphasis on enhancing the ability of machines to automatically find and use data, in ways that also happen to support reuse by humans. To accomplish this, \ac{fair} focuses on metadata in order to make data
\begin{itemize}
    \item Findable, so that metadata and data are findable for both humans and computers (e.g., persistent identifiers, rich metadata, indexing/registering);
    \item Accessible, so that users know how to access the data (e.g., clear access protocols, being able to access metadata even if the data have disappeared);
    \item Interoperable, so that data can be used by applications and workflows for analysis, storage, and processing (e.g., machine processable metadata using standards); and
    \item Reusable, so that data reuse is optimized via comprehensive, well-described metadata (e.g., metadata standards, data usage licenses, provenance).
\end{itemize}

Since that original publication, \ac{fair}'s focus on metadata has found broad application beyond data to research software~\cite{hasselbring2019}, computational workflows~\cite{goble_fair_2020}, open hardware~\cite{miljkovic2022}, \ac{ai} and \ac{ml} models~\cite{huerta2023}, and even facilities and instruments~\cite{johnson2024}. In fact, in 2021, \ac{ornl} performed some of the earliest research on applying \ac{fair} to workflows~\cite{wolf2021}, and \ac{olcf} hosted a lab-wide workshop on the topic~\cite{caw2021}. \ac{olcf} was also instrumental in recent efforts to enumerate the \ac{fair} Principles for computational workflows formally~\cite{wilkinson2025}, following in the same format and even the same journal that previously published the principles for data~\cite{wilkinson2016} and research software~\cite{barker2022}.

Following the \ac{fair} Principles helps make research artifacts more easily discoverable, shareable, and reusable. These principles improve collaboration, transparency, and efficiency in scientific research by ensuring that artifacts like code and data are well-documented, allowing them to be reused effectively across different studies and disciplines both by humans and machines. There may be challenges in implementing the \ac{fair} Principles~\cite{wilkinson_f_2022}, but they do play an important role in facilitating reproducibility, accelerating discoveries, and supporting long-term stewardship of research artifacts. Their adoption helps researchers maximize the impact of their artifacts while fostering open science and innovation.

\subsection{FAIR COMMUNITIES}

The \ac{fair} Principles focus on metadata practices to be implemented by vaguely defined ``communities''; the various publications have all deliberately left this up to interpretation. In practice, these communities tend to gather by domain (e.g. bioinformatics, geosciences, and agriculture). Such efforts have improved sharing and reusing artifacts within their respective disciplines, but these domain-based communities can end up functioning as silos. This hampers sharing and reusing artifacts as well as the adoption of the \ac{fair} Principles themselves across computational research areas.

For example, the bioinformatics community has made significant strides in developing \ac{fair} repositories such as the European Nucleotide Archive~\cite{yuan2023}. Similarly, the geoscience community has developed metadata-rich repositories such as \ac{eosdis}~\cite{ramapriyan2020}, which provides comprehensive access to satellite and remote sensing data while supporting research in areas such as climate change, natural disasters, and ecosystem monitoring. However, interoperability between these repositories remains limited, reducing the potential for cross-domain collaboration.

\subsection{CHALLENGES AT HPC CENTERS}

\ac{hpc} centers present additional challenges not present in commodity computing environments that must be addressed when implementing \ac{fair} Principles \cite{wilkinson2025-3}. In order to provide resources to users who require greater scale to ``get science done'', \ac{hpc} centers frequently deploy infrastructure with exotic hardware architectures, cutting-edge software environments, and stricter security measures as compared with users' own resources~\cite{antypas2021}. Variability in system architectures, ranging from  CPU-based clusters to GPU- and TPU-accelerated platforms, complicates software portability and reproducibility. Furthermore, interdisciplinary and inter-project collaboration within and across \ac{hpc} centers remains challenging due to security constraints, data governance policies, and technical barriers. As a result, users often create, configure, and customize digital artifacts in ways that are specialized for the unique infrastructure at a given \ac{hpc} center. This means that the research artifacts produced at \ac{hpc} centers have limited potential for reuse by others unless they are shared in ways that are discoverable by other users of those \ac{hpc} centers.

\subsection{EXISTING US DOE EFFORTS}

Several efforts in the \ac{us} \ac{doe} share the common goal of advancing scientific research by providing cutting-edge computational, data, and networking resources to support interdisciplinary collaboration: \ac{iri}~\cite{miller2023}, \ac{fasst} Framework~\cite{fasst2025}, and \ac{nairr}~\cite{nairr2025}. These initiatives focus on improving access to \ac{hpc}, data management, and security frameworks to foster cross-disciplinary research and enable more efficient, data-driven scientific discoveries. They emphasize the importance of open science principles and \ac{fair}, ensuring that research data and tools are accessible and reusable. These efforts are more concerned with enabling the autonomous laboratories of the future, however, rather than the human users of \ac{hpc} facilities.

\subsection{EOSC-LIFE FAIR WORKFLOWS COLLABORATORY}

The EOSC-Life \ac{fair} Workflows Collaboratory is a European initiative aimed at enhancing the FAIRness of scientific workflows in life sciences~\cite{goble_implementing_2021,wilkinson2025-2}. It provides a collaborative platform, including WorkflowHub~\cite{gustafsson2025}, that enables researchers to share, develop, and execute workflows across various life science domains, ensuring that data, tools, and computational resources are well-documented and standardized for broader use. It leverages the \ac{eosc} to foster collaboration between institutions and projects, promote transparency, and improve the efficiency of research through the integration of diverse data sources and computational tools. The initiative is crucial for supporting reproducibility, advancing data-driven discoveries, and accelerating innovation in European life sciences.

In this report, we argue that users could increase impactful scientific output if \ac{hpc} centers enhanced their unique environments with a FAIR-based strategy similar to the one pursued by EOSC-Life. We observe that a focus on \ac{fair} workflow components may be more impactful than focusing on \ac{fair} for entire workflows. We also propose a general architecture and discuss how to encourage adoption by users.

\section{FAIR ECOSYSTEM ARCHITECTURES}
\label{sec:ecosystems}

\begin{figure}[!ht]
\centering
    \includegraphics[width=\columnwidth]{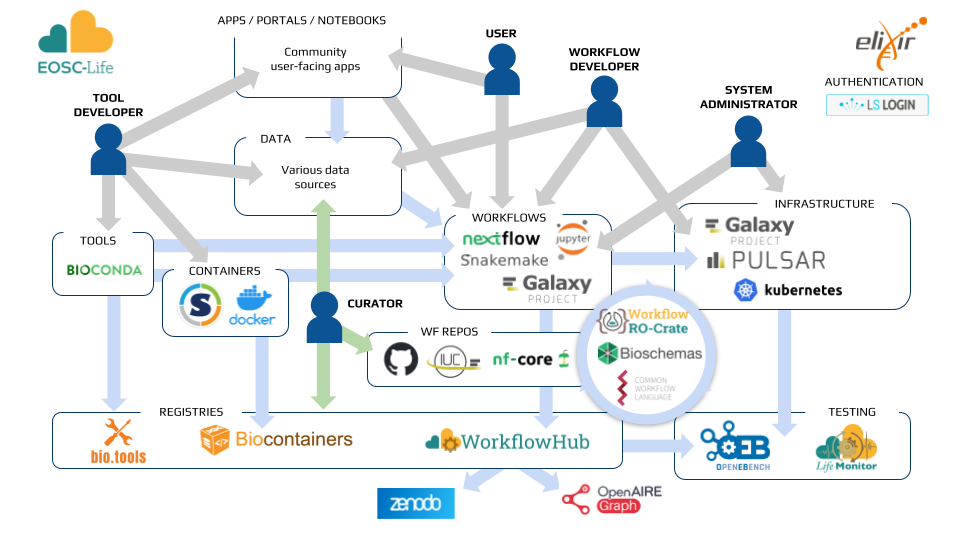}
    \caption{\label{fig:eosc-life-collaboratory}The EOSC-Life FAIR Workflows Collaboratory~\cite{wilkinson2025-2}.}
\end{figure}

Constructing a \ac{fair} ecosystem is an ambitious undertaking because such an ecosystem has many moving parts, including data, software, hardware, services, and most importantly, users. Luckily, there is an existing ecosystem for us to analyze to abstract some of the general design patterns to inform our own designs, which are detailed in Section~\ref{sec:olcf}. The EOSC-Life FAIR Workflows Collaboratory provides an example for constructing an ecosystem that incorporates \ac{fair} ``all the way down'' to aid domain scientists to produce impactful scientific artifacts that are shareable, discoverable, and reusable. A diagram depicting many of the moving parts is shown in Figure~\ref{fig:eosc-life-collaboratory}.

Although the EOSC-Life Collaboratory was constructed specifically to meet the needs of European life scientists, it does accurately portray many of the complications shared by domain scientists in general and, we argue, those faced by the users of \ac{hpc} facilities. There are some key differences, however. The EOSC-Life Collaboratory focuses mainly on the practice of open science, and life scientists try to maintain infrastructure for as long as possible to avoid breaking the pipelines they have constructed. This results in a stronger focus on applying \ac{fair} to workflows as a whole.

We argue that \ac{fair} ecosystems for \ac{hpc} centers are better served by focusing on the sharing, discovery, and reuse of \textbf{workflow components rather than entire workflows}.
This is partly because \ac{hpc} workflows are often short-lived due to being customized for \ac{hpc} resources with short design lifetimes (due to the rapid pace of hardware and software innovation). For example, the flagship machines at centers like \ac{olcf} are designed for a lifespan of 5 years; after that, executing the same workflows on a machine's successor is almost guaranteed to fail. Having witnessed this process repeatedly, we have noticed that workflow components themselves are less likely to break, but they will have to be re-plumbed to port the workflow to the new machine. Another reason for focusing on components is that, in our experience, it is extremely rare to locate an existing off-the-shelf workflow that exactly matches your needs. Users at \ac{hpc} centers come from a variety of scientific disciplines, so it is unlikely that any particular workflow will match the needs of other users exactly. Because these users are all solving challenges in trying to scale using the same \ac{hpc} resources, however, it is much more likely that solutions to separable pieces of problems will be reusable across disciplines, supporting a component-based approach.

\begin{figure}[!ht]
    \centering
    \includegraphics[width=\columnwidth]{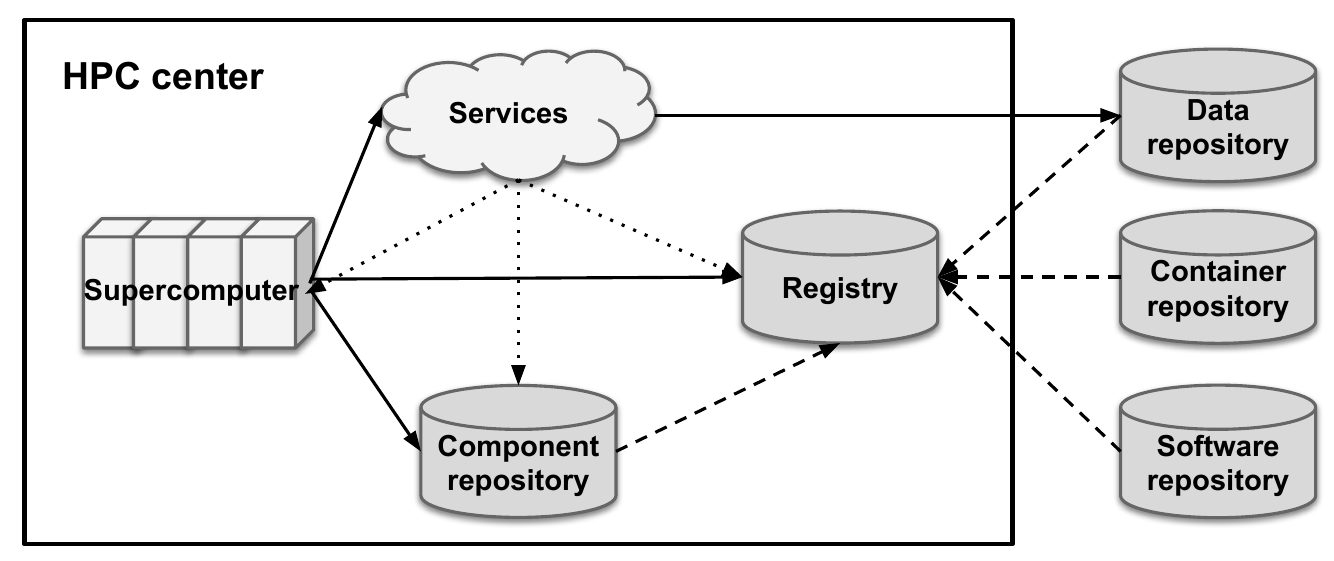}
    \caption{\label{fig:diagram}Illustrated examples of a FAIR ecosystem in action at an HPC center. Dashed lines (\textendash\textendash) show the registry's records tracking components wherever they are. Dotted lines ($\cdots$) show services in an on-prem cloud monitoring the registry and component repository for updates and providing support such as databases to supercomputer jobs. Solid lines (\textemdash) show that the supercomputer can consult the registry to locate components and request the service infrastructure to retrieve external artifacts if needed~\cite{wilkinson2025-3}.}
\end{figure}

For these reasons we propose to focus on applying \ac{fair} to workflow components, making sure they are well-described by rich metadata. This strategy requires a combination of repositories, registries, computing infrastructure, authentication/authorization, and metadata standards. This also allows us to abstract many of the details from Figure~\ref{fig:eosc-life-collaboratory} into the simpler Figure~\ref{fig:diagram}, which depicts an abstract \ac{fair} ecosystem at an \ac{hpc} center.

\subsection{REPOSITORIES}

Repositories, roughly speaking, are places where \textit{things} are stored; this is in contrast to registries (see below), which are where \textit{records} of things are stored. Registries and repositories provide curation and best practices for recording workflows \cite{wcs2022}. Workflow components are what workflows are made of, which can include various forms of data (e.g. simulation output, trained \ac{ai}/\ac{ml} models, and provenance) and software (e.g. scripts, programs, and computational workflows). Having a robust repository system aligns with \ac{iri}-style distributed, collaborative science and its aims for creating a seamless, interoperable research infrastructure across \ac{doe} labs.

The \ac{fair} Principles do not imply openness, but we also recommend following open science principles~\cite{nasem2018} when possible by enabling integration with externally hosted public repositories (e.g. Conda\footnote{\url{https://anaconda.org/anaconda/conda}}, Spack\footnote{\url{https://spack.io/}}, Docker Hub\footnote{\url{https://hub.docker.com/}}, TensorFlow Hub\footnote{\url{https://www.tensorflow.org/hub}}). Security or scale concerns may not allow for open dissemination or integration, however. \ac{hpc} centers typically handle their own user identity management, and those operated by \ac{us} Government entities must conform to well-specified security policies. Scale of workflow components tends to be more a technical issue; centers typically address these through data transport middleware designed for large scale (e.g. Globus~\cite{globus}) and specialized repositories (e.g. DataFed~\cite{stansberry2019}) which can encapsulate both data transport and storage capabilities. These concerns can also take different forms for commercial cloud compute providers than at leadership-class \ac{hpc} centers where the ability to scale in compute and data is a goal in itself.

\subsection{REGISTRIES}

Registries, as mentioned previously, are places where records of artifacts are stored, and they can be distinct from the repositories that contain the artifacts described by the records. A well-designed registry such as WorkflowHub~\cite{gustafsson2025} is critically important for enabling a \ac{fair} ecosystem anywhere, but particularly as illustrated in~\ref{fig:diagram}, they are crucial. Registries complement repositories by maintaining metadata records for workflow components, ensuring the discoverability and interoperability of research artifacts via metadata and \ac{pids} such as \ac{dois}. In this case, the registry stores metadata about the artifacts as well as where they are stored, allowing users to search for artifacts relevant to their needs across repositories both internal and external to the \ac{hpc} center.

\subsection{COMPUTING INFRASTRUCTURE}

Computing infrastructure, although not digital, must still be described by rich metadata. Reproducibility and reusability rely on preservation and discovery of the computing contexts in which workflows are executed. It is therefore important that metadata about the resources and infrastructure provided by an \ac{hpc} center be made available and exportable in useful forms. Tools such as FlowCept~\cite{souza2023} help to capture provenance about execution and environment that should be stored and linked to the workflow components. Reproducing execution context is a critical enabler for the seamless cross-center data exchange and workflow portability that \ac{iri} needs.

Following the model of the EOSC-Life Collaboratory~\cite{goble_implementing_2021} requires infrastructure for \emph{service} execution, separate from but adjacent to \ac{hpc} supercomputing resources. Infrastructure provided by Kubernetes\footnote{\url{https://kubernetes.io/}}- and OpenStack\footnote{\url{https://www.openstack.org/}}-based platforms, for example, enables persistent services for automatically testing workflow component viability as well as continuously indexing and updating registries in response to artifact updates in repositories. Commonly available services running on instances of these platforms also reduce the need for workflows to deploy their own versions (e.g., a key-value store). Validation and benchmarking services help align with \ac{nairr}, ensuring that \ac{ai}/\ac{ml} models trained on \ac{hpc} resources remain \ac{fair}-compliant and reusable even as the \ac{fair} Principles are updated in the future.

\subsection{AUTHENTICATION AND AUTHORIZATION}

Authentication and authorization via \ac{rbac} mechanisms protect sensitive workflow components while maintaining accessibility for authorized users, ensuring that only the right users can access and use certain components. This is an underappreciated aspect of the ``A'' in \ac{fair}, which stands for Accessibility. Although \ac{fair} is frequently used in conjunction with open science principles, as we noted above \ac{fair} does not strictly imply openness, which is a fact that even seasoned \ac{fair} practitioners can forget.

Rich metadata about workflow components should include information about who can use the components, and this usually falls under the ``R'' for Reusability. When that metadata is made available in the registry, it can feed appropriate authentication and authorization procedures that fall under the ``A'' in \ac{fair}. These problems are not unique to \ac{hpc} environments, but they must still be solved in \ac{fair} ecosystems for \ac{hpc}. Authentication and authorization mechanisms, including federated identity management solutions such as OAuth\footnote{\url{https://oauth.net/2/}} and OpenID Connect\footnote{\url{https://openid.net/developers/how-connect-works/}}, facilitate the secure cross-institutional collaboration necessary for distributed, collaborative computational science.

One of the other important requirements is that the workflows
should not hardcode any credentials either through workflow
specification files or programming; credentials must be kept
separate \cite{vonlaszewski2025}. This problem affects the reproducibility and reusability of workflows and their components.

\subsection{METADATA STANDARDS AND INTERCHANGE FORMATS}

Metadata standards and interchange formats are the glue that holds the whole ecosystem together; they play a crucial role in enabling \ac{fair} at all scales \cite{wilkinson2025-3}. The EOSC-Life Collaboratory's WorkflowHub\footnote{\url{https://workflowhub.eu/}}~\cite{gustafsson2025} uses a rich metadata framework for findability with inter-registration integration with sister registries for containers and tools using \ac{apis} and shared \ac{pids}. Tools and workflows are marked up with metadata standards from Bioschemas~\cite{gray2017}, encouraging the enrichment of metadata by workflow and tool developers, including licensing and provenance metadata. Despite recent work to establish a standard \ac{wms} terminology~\cite{suter2025}, a number of challenges remain, so we recommend abstract \ac{cwl}~\cite{crusoe2022} as a canonical representation of the workflow. WorkflowHub imports and exports the exchange standard Workflow-RO-Crates~\cite{leo2024}, facilitating automatic builds for each of its entries, regardless of whether the management system natively supports it. The architecture here follows that example closely not only because it helps make workflows ``good''~\cite{wcs2024} but also because this architecture seeks to integrate directly with WorkflowHub, in alignment with \ac{fasst}’s efforts to streamline scientific workflows and training mechanisms.

\section{PROPOSED ARCHITECTURE FOR OLCF}
\label{sec:olcf}
%

Having provided an overview of a \ac{fair} ecosystem's architecture in Section~\ref{sec:ecosystems}, we now propose concrete details for implementing such an ecosystem for the users of \ac{olcf}. We model this implementation after the example of the EOSC-Life \ac{fair} Workflows Collaboratory, taking into account that there are different constraints and needs of \ac{olcf}'s users.

One major difference for \ac{olcf} is that its users may want or require higher levels of security and privacy than the open science model used by EOSC-Life. For example, users may wish to keep their code or datasets private for a time (e.g. embargo) until results have been fully analyzed and published. Users might also be required to keep their research artifacts secret (e.g. export control).

\begin{figure}[!ht]
    \centering
    \includegraphics[width=\columnwidth]{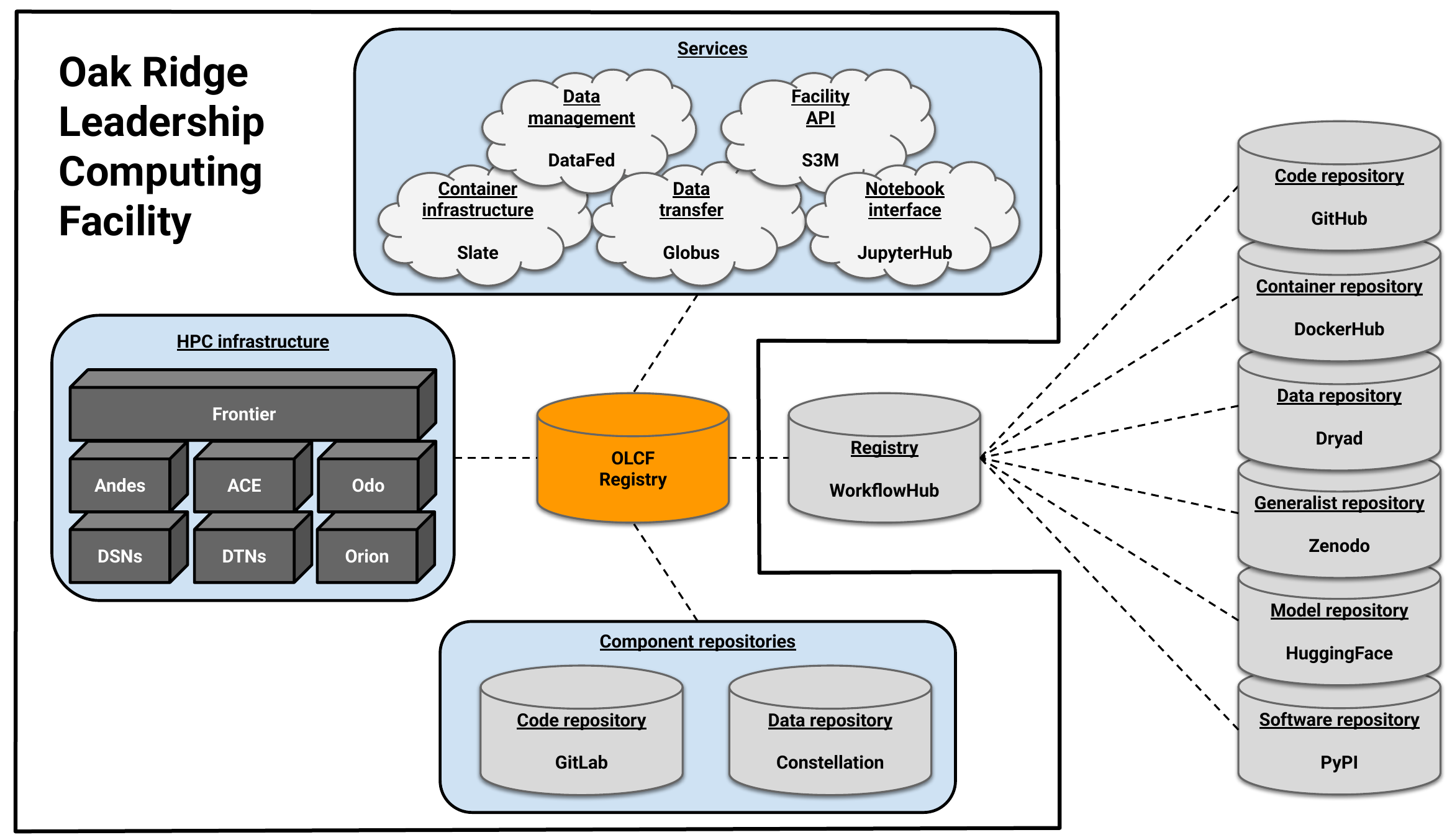}
    \caption{\label{fig:concrete}Illustrated example of emerging FAIR ecosystem at OLCF. This is a concrete version of Figure~\ref{fig:diagram} in which specific infrastructure and services at OLCF are named explicitly. Note that ``OLCF Registry'' (orange center object) does not exist yet; this is a key difference between the two figures.}%
\end{figure}

Recall that in Section~\ref{sec:ecosystems}, we abstracted the details away from Figure~\ref{fig:eosc-life-collaboratory} in order to present the general patterns in Figure~\ref{fig:diagram}. We now add details specific to \ac{olcf} in order to produce Figure~\ref{fig:concrete}. In this section, we propose implementations for each of the main ``parts'' which remain: repositories, registries, infrastructure, and services.

\subsection{REPOSITORIES}

As outlined in Section~\ref{sec:ecosystems}, repositories are places where \textit{things} are stored, and we have argued that the things of interest for an \ac{hpc} center are workflows components, rather than whole workflows. A workflow itself may be conceptualized as the formal specification of the data flow and execution control between executable components, expected datasets, and parameter files. Workflow components can be executables and data. Examples of executable components include scripts, code, tools, containers, or workflows themselves remotely or locally executed, and native or third party; an example data component could be a reference dataset such as the Human Reference Genome~\cite{wilkinson2025}.

We argue that the individual workflow components are more likely to be useful across scientific domains than entire workflows, which means that a focus on storing workflow components is more likely to benefit the users of \ac{olcf}, who represent a diverse collection of scientific disciplines. As shown in Figure~\ref{fig:eosc-life-collaboratory}, the EOSC-Life Collaboratory incorporates a number of different repositories for different types of components, most of which are for open science. These include:
\begin{itemize}
    \item code repositories like GitHub\footnote{\url{https://github.com}},
    \item container repositories like DockerHub,
    \item software repositories like BioConda\footnote{\url{https://bioconda.github.io/}},
    \item workflow repositories which are \ac{wms}-specific like Intergalactic Utilities Commission\footnote{\url{https://galaxyproject.org/iuc/}} for Galaxy\footnote{\url{https://galaxyproject.org/}} and nf-core\footnote{\url{https://nf-co.re/}} for Nextflow\footnote{\url{https://www.nextflow.io/}}, and
    \item generalist repositories such as Zenodo\footnote{\url{https://zenodo.org/}} where nearly any type of artifact may be stored.
\end{itemize}

As mentioned previously, not all users of \ac{olcf} perform open science. For those who do perform open science, all the same component repositories used by EOSC-Life are readily available to store their components. Figure~\ref{fig:concrete} depicts a solution which satisfies the additional constraints by providing private component repositories inside \ac{olcf} for code and data using solutions already present at \ac{olcf}.

\textbf{Constellation}\footnote{\url{https://doi.ccs.ornl.gov}}~\cite{vazhkudai2016} is a scalable, flexible, and extensible data infrastructure developed and hosted by \ac{olcf} to support scientific discovery through data sharing, reproducibility, and collaboration. Among other features, it provides a DOI minting service, which would allow it to serve as a generalist repository for published research artifacts. Constellation is typically used for datasets, but there is no reason why it cannot be used to host compiled executables, pre-built containers, trained models, and the provenance data from specific workflow runs. In this way, Constellation could serve as a foundational platform for hosting arbitrary artifacts -- including workflow components -- across scientific domains, offering standardized metadata, persistent identifiers, and APIs that facilitate discovery, access, and reuse by both humans and machines, thus bridging siloed research communities and supporting open science at scale.

\ac{olcf} already provides \textbf{GitLab}\footnote{\url{https://gitlab.ccs.ornl.gov/}} to its users for collaborating on code development, as a local alternative to GitHub. This is a valid code repository for storing source code, scripts, tools, workflow specifications and small datasets and parameter files for testing and demonstration purposes.

Additionally, \textbf{DataFed}\footnote{\url{https://datafed.ornl.gov}}~\cite{stansberry2019} is a lightweight, distributed scientific data management system developed by \ac{olcf} that spans a federation of storage systems within a loosely-coupled network of scientific facilities. In Figure~\ref{fig:concrete}, DataFed is shown as a service, but it facilitates ``data development'' in a sense. As such, it bears mentioning here as a data repository for data that have not yet reached the publication stage but which \ac{olcf} users might wish to share with each other.

\subsection{REGISTRIES}

As outlined in Section~\ref{sec:ecosystems}, registries are places where \textit{records} of artifacts (``things'') are stored. Because \ac{fair} is essentially about metadata and persistent identifiers, registries are the ``secret sauce'' of a \ac{fair} ecosystem. This is because when users register their artifacts in a registry designed with the \ac{fair} Principles in mind, their artifacts immediately become more Findable, Accessible, and Reusable. In fact, the sub-principle F4 from the original paper is that ``(meta)data are registered or indexed in a searchable resource''~\cite{wilkinson2016}.

The EOSC-Life FAIR Workflows Collaboratory incorporates several different registries to enable searches across the different repositories that store their artifacts, as shown in Figure~\ref{fig:eosc-life-collaboratory}. The registries were all developed specifically for the \ac{elixir} project, and they include:
\begin{itemize}
    \item bio.tools\footnote{\url{https://bio.tools/about}}, a curated online resource that provides standardized information and metadata about bioinformatics tools and software to support research and tool discovery~\cite{ison2015,ison2019},
    \item BioContainers\footnote{\url{https://biocontainers.pro/}}, an open-source platform that provides standardized, containerized bioinformatics software to facilitate reproducible research and easy tool deployment~\cite{daveigaleprevost2017}, and
    \item WorkflowHub, a domain-agnostic, FAIR-compliant registry for describing, sharing and publishing scientific computational workflows~\cite{gustafsson2025}.
\end{itemize}

Unlike the previous case for repositories, \textbf{OLCF currently has no corresponding registries}, which is why the ``\ac{olcf} registry'' is highlighted  in orange in Figure~\ref{fig:concrete}. As mentioned earlier, the \ac{fair} Principles call for a registry not by name, but by description. For this reason, any strategy for constructing a \ac{fair} ecosystem for \ac{olcf} users must necessarily include a registry. 

Currently, the only option for \ac{olcf} users to register workflow components is to use WorkflowHub, which will mainly serve the needs of the users who work in open science. WorkflowHub is free to use, and it is publicly available as a website which also provides an API. Access to artifacts themselves may be controlled so that they remain ``private'', but WorkflowHub still shows that the artifacts exist. This is by design; it is a more advanced consequence of the \ac{fair} principles. As mentioned previously, however, \ac{olcf} users sometimes have reasons that prevent making their artifacts fully open to the entire world immediately. Additionally, the features needed to search WorkflowHub for workflow components that can execute on the unique infrastructure at \ac{olcf} is still in the ``feature request'' stage. For these reasons, \ac{olcf} needs to deploy its own registry in-house to meet the needs of its users.

A few options for standing up a registry immediately come to mind.

There are at least three options for how to do this:
\begin{enumerate}
    \item leverage DataFed as a ``metadata repository'', either directly or by developing a user interface backed by DataFed as storage;
    \item develop our own registry completely from scratch, which is usually the most labor-intensive option; or
    \item deploy our own instance of WorkflowHub inside \ac{olcf}, most likely as a containerized service running on Slate.
\end{enumerate}

More information may be required in order to make a full recommendation, however.

\subsection{COMPUTING INFRASTRUCTURE}

As mentioned previously, even though computing infrastructure is physical rather than digital, it must still be described by rich metadata. Reproducibility and reusability rely on preservation and discovery of the computing contexts in which workflows are executed. It is therefore important that metadata about the resources and infrastructure provided by an \ac{hpc} center be made available and exportable in useful forms. In fact, the supercomputers at \ac{hpc} facilities should be treated as scientific instruments, to which applying the \ac{fair} Principles has recently emerged as an active area of research~\cite{johnson2024, julian2024}.

The model of the EOSC-Life Collaboratory separates service execution from \ac{hpc} supercomputing resources, as shown in Figure~\ref{fig:eosc-life-collaboratory}. \ac{olcf} is very well positioned to emulate this model already. As shown in Figure~\ref{fig:concrete}, \ac{olcf} provides OpenShift\footnote{\url{https://docs.openshift.com/}} on \textbf{Slate} to provide users the capability to deploy their own services (e.g. databases, workflow management systems) adjacent to the \textbf{Frontier} supercomputer that persist beyond the end of individual Frontier batch jobs. Thus, \ac{olcf} already have the capability right now for deploying services that automatically test workflow component viability, for example, just like the EOSC-Life Collaboratory. One key difference is that because \ac{olcf} provides self-management capabilities to its users, it does not provide shared common services (e.g. a multi-user key-value store).

One way in which \ac{olcf} exceeds the model of the EOSC-Life Collaboratory is that it has been developing the \textbf{Secure Scientific Service Mesh (S3M)}, which provides API-driven infrastructure to accelerate scientific discovery through automated research workflows~\cite{skluzacek2025}. It integrates near real-time streaming capabilities, intelligent workflow orchestration, and fine-grained authorization within a service mesh architecture to enable secure and flexible programmatic access to \ac{hpc} resources. By providing an API that allows intelligent agents and experimental facilities to provision resources dynamically and execute complex workflows, S3M significantly reduces traditional barriers between \ac{olcf}'s users, its computational resources, and other experimental facilities.


\subsection{AUTHENTICATION AND AUTHORIZATION}

The \ac{fair} Principles advocate for robust authentication and authorization procedures as a crucial element of the ``Accessible'' principle. This is particularly true for sensitive data, \ac{ai} models trained on sensitive data, software for handling sensitive data, and other kinds of workflow components that cannot be made openly available. Ensuring that data is accessible to the right users under the right conditions aligns with \ac{fair}'s goal of maximizing the value of artifacts while respecting privacy and security concerns.

Despite the fact that much of the research conducted using the EOSC-Life Collaboratory is open, the Collaboratory does provide Life Science Login\footnote{\url{https://lifescience-ri.eu/ls-login/}} to ensure that only the right users can access artifacts and resources under the right conditions.

\ac{olcf} provides multiple security enclaves which are completely separated from each other, even at the account level. From a user's perspective, each of these enclaves can still be considered its own \ac{fair} ecosystem. There is nothing about \ac{fair} which implies openness; accessibility in the \ac{fair} sense refers to the fact that everyone in a given ``community'' can access artifacts and resources over protocols that provide capabilities for authentication and authorization. Some of the ways in which \ac{fair} practices such as persistent identifiers and metadata figure into security for a facility would likely be more useful to facility staff such as system administrators, rather than users. System administrators are still part of the ecosystem, as shown in Figure~\ref{fig:eosc-life-collaboratory}, but those discussions are beyond the scope of this report.

We do not suggest any changes from current design at this time that would be visible from the perspective of an \ac{olcf} user.


\subsection{METADATA STANDARDS AND INTERCHANGE FORMATS}


Finally, metadata standards and interchange formats hold the entire ecosystem together through their importance in enabling the ``Interoperable'' part of the \ac{fair} Principles. It is challenging to provide recommendations for \ac{olcf} here, however, for social reasons that are discussed in Section~\ref{sec:adoption}. For now, we recommend following many of the practices established by the EOSC-Life FAIR Workflows Collaboratory.

In particular, the WorkflowHub registry~\cite{gustafsson2025} its in the middle of everything, as shown in Figure~\ref{fig:eosc-life-collaboratory}, and it is enabled by judicious choices for metadata standards like Bioschemas~\cite{gray2017} and interchange formats like RO-Crate~\cite{soiland-reyes2022}. WorkflowHub uses the Global Alliance for Genomics and Health (GA4GH) Tools Registry Service (TRS) API~\cite{rehm2021} to integrate with sister registries; \ac{olcf} will likely wish to use the same framework so that it can integrate its internal registry of the future with WorkflowHub, too.

Although WorkflowHub supports workflows of all kinds from all workflow management systems, some systems provide extra features through deeper integrations. Although there is no universal way to build a workflow, we do recommend abstract \ac{cwl}~\cite{crusoe2022} for a canonical representation of the workflow because of its deeper integrations with WorkflowHub. \ac{cwl} is also supported by a number of workflow management systems that target very different execution platforms (e.g. cloud, Java Virtual Machine, containers), which provides hope that workflows composed in \ac{cwl} might have a longer lifespan. When it is time to package workflows and their components, WorkflowHub imports and exports using exchange formats RO-Crate~\cite{soiland-reyes2022} and Workflow-RO-Crate~\cite{leo2024}, the latter of which facilitates automatic builds for each of its entries, regardless of whether the workflow management system natively supports it.

As mentioned previously, it is difficult to make recommendations on behalf of the entire community of \ac{olcf} users for social reasons, namely that \ac{fair} delegates exact implementation details to ``community standards'' that should be chosen by communities themselves. We discuss these social issues in the next section.

\section{ENCOURAGING ADOPTION OF FAIR AT OLCF}
\label{sec:adoption}

The \ac{olcf} serves a broad and diverse scientific community of users and provides access to some of the world's most powerful \ac{hpc} resources. This community generates valuable datasets, \ac{ai} models, software, and other artifacts across a wide range of disciplines, from climate modeling and materials science to molecular biology and astrophysics. To accomplish this, the users frequently solve the same problems individually for which solutions might, in an ideal world, already exist, even for the unique systems at \ac{olcf}. Such problems include adapting file reads and writes for the shared filesystems, tuning parameters to optimize parallel performance, connecting persistent services from Slate to Frontier, and many other small tasks. Solving these problems produces artifacts that could be reused and improved upon by others, maximizing impact beyond even originally intended purposes. The \ac{fair} Principles provide guidelines that help make artifacts shareable, discoverable, and reuseable by others, but their adoption by \ac{olcf}'s users is a problem of its own.

Despite the clear benefits of \ac{fair} practices, their widespread implementation within the OLCF user community remains inconsistent. This is due in part to deeply rooted research norms, varied domain-specific data practices, and the perceived cost or complexity of adopting new standards. Driving cultural change in such an environment requires not only policy guidance and technical support, but also a systematic approach to shifting community behaviors and incentives.

\begin{figure}[!ht]
    \centering
    \includegraphics[width=0.75\columnwidth]{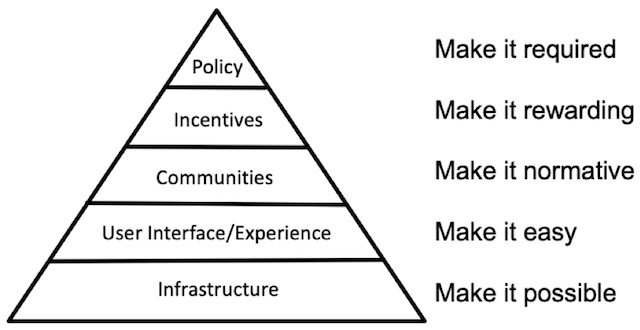}
    \caption{\label{fig:nosek} Strategy for Culture Change. Image credit: \url{https://www.cos.io/blog/strategy-for-culture-change}.}
\end{figure}

In this context, the ``Strategy for Culture Change'' proposed by Brian Nosek in a 2019 blog post\footnote{\url{https://www.cos.io/blog/strategy-for-culture-change}} provides a useful framework. Nosek suggests that sustainable cultural transformation can be achieved by altering the default behaviors in a system through three coordinated levers: infrastructure, incentives, and norms. Applying this framework to the OLCF community offers a pathway to embed the \ac{fair} Principles into standard research workflows gradually, making the \ac{fair} choice not only possible, but easy and rewarding.

The diagram shown in Figure~\ref{fig:nosek} illustrates an approach that works from the bottom up, showing that for us to encourage adoption of \ac{fair} among \ac{olcf} users, we must first make adoption possible, especially at the infrastructure level. As we showed in Section~\ref{sec:olcf}, \ac{olcf} already possesses most of the pieces to construct a \ac{fair} ecosystem. The most glaring hole that is missing is a registry for tracking components. \textbf{OLCF needs a registry to make a FAIR ecosystem possible.}

After making adoption possible, Nosek recommends that \ac{olcf} make adoption easy. More often than not, \ac{olcf} users are highly technically skilled in addition to being leaders in their scientific fields. It is a lot to ask these users to become \ac{fair} experts, too! We can ``make it easy'' by providing simple user interfaces to walk them through what would otherwise seem like very complicated processes for registering and sharing their artifacts, or for finding relevant existing artifacts they can reuse. Some efforts are already underway, such as the \textbf{FlowCept} tool~\cite{souza2023} for capturing provenance about a workflow's execution and environment; we should also provide interfaces for storing and linking to the workflow components themselves. The API for \textbf{S3M} is a user interface that is designed to simplify some very complicated systems on behalf of the user. In both cases, adopting \ac{fair} practices is already possible, but \ac{olcf} is providing tools and services that make adoption significantly easier.

Next, Figure~\ref{fig:nosek} shows that \ac{olcf} should make \ac{fair} adoption normative, which requires psychology beyond the scope of this report to fully explain. The example given by Nosek for shifting norms is to \textbf{make the desired behaviors visible}, such as when journals adopt badges to acknowledge authors who preregister their studies and share their data and materials. Those awarded badges become signals to other researchers as they read the articles both that their colleagues do these behaviors  (descriptive norm setting) and that the journal values these behaviors (prescriptive norm setting). At \ac{olcf}, we might spotlight certain projects in newsletters or at the monthly user group meetings to increase visibility for the desired behaviors. Visibility is critical for accelerating adoption among those who are willing but who have not yet adopted.

For those users who are not quite as willing, the remaining two steps of the strategy attempt to make adoption rewarding before ultimately making adoption required. Colloquially, this is also known as the ``carrot and stick'' metaphor\footnote{\url{https://en.wikipedia.org/wiki/Carrot_and_stick}}, which evokes imagery of leading a horse either towards a positive incentive (eating a tasty carrot) or away from a negative consequence (being forcefully thrashed with a stick).

Positive incentives for users include the creation of reward systems that recognize and incentivize researchers for sharing their components. This can be achieved through citation and attribution mechanisms which formally acknowledge the contributions of those who make their computational resources, data, or code available for public use. At an \ac{hpc} center like \ac{olcf}, however, there are other currencies available, too, such as increased allocations and job priority. Allocations (measured in node-hours) and job priority (measured in wait time) both represent ways to \textbf{reward time back to users who actively work to save time for other users}. It will be important to balance such rewards against potential resulting disincentives, such as users' equating reuse of shared components with reduction in resources available to themselves.

Rather than utilizing ``negative incentives'', however, \ac{olcf} would most likely enact and enforce policies to make adoption required among the final users who have still held out. Receiving allocations of resources at \ac{olcf} could be conditioned on \ac{fair} compliance and participation in the workflow component and research artifact-sharing ecosystem. Currently, allocations are frequently awarded at no cost to researchers or are governed by proposal processes funded by \ac{us} taxpayers. Artifacts resulting from that usage should (perhaps after some embargo period to protect intellectual investment pending publication) be freely available and reusable. Researchers will want assurances that they will be appropriately credited for their efforts, and so implementation of such mandates will require careful consideration.

The incentives for \ac{olcf} to construct a \ac{fair} ecosystem and encourage its adoption are important, too. Not least among these incentives is to uphold the mission of \ac{olcf} itself, which is to accelerate scientific discovery and engineering progress by providing world-leading computational performance and advanced data infrastructure. The \ac{fair} Principles maximize the value and impact of research artifacts~\cite{wilkinson2025}, which is critically important for artifacts that are the products of hundreds of thousands of node-hours on a Top500\footnote{\url{https://top500.org}} machine like Frontier~\cite{atchley2023}. It is prohibitively expensive to reproduce such artifacts, when it is possible to reproduce them at all without access to a leadership-class supercomputer. Frontier is a time-sensitive scientific instrument with a finite lifespan; avoiding recomputing results not only speeds time to discovery for subsequent studies, but it also frees the machine to continue crunching numbers on behalf of new artifacts that could lead to additional discoveries. In short, the benefits of adopting \ac{fair} allow Frontier to produce more results that can lead to more scientific discovery, thereby increasing the overall scientific impact of Frontier and of \ac{olcf}.


\section{FUTURE DIRECTIONS}
\label{sec:future}
%




Having discussed in previous sections why and how \ac{fair} relates to \ac{hpc} centers in general and \ac{olcf} in particular, we now similarly connect and relate \ac{fair} with the recently announced \ac{asc}.

At the time of this writing, not much is publicly known about the \ac{asc} beyond what has been revealed by the One Big Beautiful Bill \cite{big-bill}, which defines it as ``a system of United States government, academic, and private sector programs and infrastructures utilizing cloud computing technologies to facilitate and support scientific research, data sharing, and computational analysis across various disciplines while ensuring compliance with applicable legal, regulatory, and privacy standards.'' A bundled summary of its enclosing section provides a little more information, however:
\begin{quotation}
    This section provides funding for partnerships between the National Laboratories and U.S. industry to organize DOE data for use in artificial intelligence and machine learning models. DOE must also initiate seed efforts for self-improving artificial intelligence models for science and engineering using this data. These models must be provided to the scientific community through a system of programs and infrastructure using cloud computing. This section also allows this data to be used to develop next-generation microelectronics.
\end{quotation}

From this information, we can deduce that the future \ac{asc} may have quite a lot in common with the \ac{eosc}, the website\footnote{\url{https://eosc.eu/eosc-about/}} for which states that ``[t]he ambition of the European Open Science Cloud, known as EOSC, is to develop a \textbf{`Web of FAIR Data and Services'} for science in Europe.'' This sentence is intended to evoke the ``Web of Linked Data'' \cite{bernerslee2006} concept introduced in 2006, as updated by the \ac{fair} Principles published 10 years later \cite{wilkinson2016}. Interestingly, the European Commission began pursuing \ac{eosc} in 2015, which now provides 10 years' worth of potentially relevant insights for constructing the \ac{asc}.

In Sections~\ref{sec:ecosystems} and \ref{sec:olcf}, we repeatedly referenced insights and patterns from the EOSC-Life FAIR Workflows Collaboratory when designing a \ac{fair} ecosystem for workflows for \ac{hpc} centers and for \ac{olcf}. Without loss of generality, these same arguments and recommendations would apply to the broader \ac{asc}, from what is currently known. The \ac{asc} does seem to place greater focus on \ac{ai} / \ac{ml} models than the EOSC-Life ecosystem shown in Figure~\ref{fig:eosc-life-collaboratory}, however, and so for that, we recommend following the work of the \ac{doe}-funded HPC-FAIR project \cite{liao2024}.

HPC-FAIR was a multi-institutional project that aimed to develop a generic \ac{hpc} data management framework to make both training data and \ac{ai} models of scientific applications \ac{fair} \cite{huerta2023}. The framework was designed to centralize \ac{hpc}-related datasets and \ac{ai} models within a unified hub and to ensure interoperability through a standardized representation and vocabulary (ontology) for both data and models. The project also implemented automated workflows for streamlined data processing, model access, and benchmarking, and it focused on optimizing data harnessing efficiency through advanced techniques like deep reuse and compression-based analytics \cite{liao2024}.

In a nutshell, the application of the \ac{fair} Principles at every level of an ecosystem -- be it inside \ac{olcf} or a much more geographically distributed system like the \ac{asc} -- is crucial for enabling autonomous, self-assembling workflows. Although \ac{llm}s have proven capable of generating performant code on demand \cite{valero-lara2024}, the value of building from vetted components has been a cornerstone of software engineering for quite some time. For any human- or machine-based system to build from components, those components will inevitably need to be Findable, Accessible, Interoperable, and Reusable, which immediately implies that many of us who lay the groundwork now \textbf{must learn FAIR or waste time reinventing it}. \ac{fair} is certain to be a critical enabler and accelerator of the construction of the \ac{asc}.

\cleardoublepage\phantomsection\addcontentsline{toc}{section}{REFERENCES}
\bibliographystyle{IEEEtranDOI}
\bibliography{references}

@inproceedings{wilkinson_f_2022,
	address = {Salt Lake City, UT, USA},
	title = {F*** workflows: when parts of {FAIR} are missing},
	copyright = {https://doi.org/10.15223/policy-029},
	isbn = {9781665461245},
	shorttitle = {F*** workflows},
	url = {https://doi.org/10.1109/eScience55777.2022.00090},
	doi = {10.1109/eScience55777.2022.00090},
	urldate = {2024-08-07},
	booktitle = {2022 {IEEE} 18th {International} {Conference} on e-{Science} (e-{Science})},
	publisher = {IEEE},
        author = {Wilkinson,  Sean R. and Eisenhauer,  Greg and Kapadia,  Anuj J. and Knight,  Kathryn and Logan,  Jeremy and Widener,  Patrick and Wolf,  Matthew},
	month = oct,
	year = {2022},
	pages = {507--512},
}

@article{wilkinson2016,
	title = {The {FAIR} {Guiding} {Principles} for scientific data management and stewardship},
	volume = {3},
	issn = {2052-4463},
	url = {https://doi.org/10.1038/sdata.2016.18},
	doi = {10.1038/sdata.2016.18},
	language = {en},
	number = {1},
	urldate = {2024-07-08},
	journal = {Scientific Data},
        author = {Wilkinson,  Mark D. and Dumontier,  Michel and Aalbersberg,  IJsbrand Jan and Appleton,  Gabrielle and Axton,  Myles and Baak,  Arie and Blomberg,  Niklas and Boiten,  Jan-Willem and da Silva Santos,  Luiz Bonino and Bourne,  Philip E. and Bouwman,  Jildau and Brookes,  Anthony J. and Clark,  Tim and Crosas,  Mercè and Dillo,  Ingrid and Dumon,  Olivier and Edmunds,  Scott and Evelo,  Chris T. and Finkers,  Richard and Gonzalez-Beltran,  Alejandra and Gray,  Alasdair J.G. and Groth,  Paul and Goble,  Carole and Grethe,  Jeffrey S. and Heringa,  Jaap and ’t Hoen,  Peter A.C and Hooft,  Rob and Kuhn,  Tobias and Kok,  Ruben and Kok,  Joost and Lusher,  Scott J. and Martone,  Maryann E. and Mons,  Albert and Packer,  Abel L. and Persson,  Bengt and Rocca-Serra,  Philippe and Roos,  Marco and van Schaik,  Rene and Sansone,  Susanna-Assunta and Schultes,  Erik and Sengstag,  Thierry and Slater,  Ted and Strawn,  George and Swertz,  Morris A. and Thompson,  Mark and van der Lei,  Johan and van Mulligen,  Erik and Velterop,  Jan and Waagmeester,  Andra and Wittenburg,  Peter and Wolstencroft,  Katherine and Zhao,  Jun and Mons,  Barend},
	month = mar,
	year = {2016},
	pages = {160018},
}

@article{wilkinson2025,
  title = {Applying the {FAIR Principles} to computational workflows},
  volume = {12},
  ISSN = {2052-4463},
  url = {https://doi.org/10.1038/s41597-025-04451-9},
  DOI = {10.1038/s41597-025-04451-9},
  number = {1},
  journal = {Scientific Data},
  publisher = {Springer Science and Business Media LLC},
  author = {Wilkinson,  Sean R. and Aloqalaa,  Meznah and Belhajjame,  Khalid and Crusoe,  Michael R. and de Paula Kinoshita,  Bruno and Gadelha,  Luiz and Garijo,  Daniel and Gustafsson,  Ove Johan Ragnar and Juty,  Nick and Kanwal,  Sehrish and Khan,  Farah Zaib and K\"{o}ster,  Johannes and Peters-von Gehlen,  Karsten and Pouchard,  Line and Rannow,  Randy K. and Soiland-Reyes,  Stian and Soranzo,  Nicola and Sufi,  Shoaib and Sun,  Ziheng and Vilne,  Baiba and Wouters,  Merridee A. and Yuen,  Denis and Goble,  Carole},
  year = {2025},
  month = feb 
}

@article{goble_fair_2020,
	title = {{FAIR} {Computational} {Workflows}},
	volume = {2},
	issn = {2641-435X},
	url = {https://doi.org/10.1162/dint\_a\_00033},
	doi = {10.1162/dint\_a\_00033},
	number = {1-2},
	journal = {Data Intelligence},
    author = {Goble,  Carole and Cohen-Boulakia,  Sarah and Soiland-Reyes,  Stian and Garijo,  Daniel and Gil,  Yolanda and Crusoe,  Michael R. and Peters,  Kristian and Schober,  Daniel},
	year = {2020},
	pages = {108--121},
}

@article{goble_implementing_2021,
	title = {Implementing {FAIR} {Digital} {Objects} in the {EOSC}-{Life} {Workflow} {Collaboratory}},
	copyright = {Creative Commons Attribution 4.0 International, Open Access},
	url = {https://doi.org/10.5281/ZENODO.4605654},
	doi = {10.5281/ZENODO.4605654},
	urldate = {2024-02-20},
        author = {Goble,  Carole and Soiland-Reyes,  Stian and Bacall,  Finn and Owen,  Stuart and Williams,  Alan and Eguinoa,  Ignacio and Droesbeke,  Bert and Leo,  Simone and Pireddu,  Luca and Rodríguez-Navas,  Laura and Fernández,  José Mª and Capella-Gutierrez,  Salvador and Ménager,  Hervé and Gr\"{u}ning,  Bj\"{o}rn and Serrano-Solano,  Beatriz and Ewels,  Philip and Coppens,  Frederik},
	month = mar,
	year = {2021},
}

@article{barker2022,
	title = {Introducing the {FAIR} {Principles} for research software},
	volume = {9},
	issn = {2052-4463},
	url = {https://doi.org/10.1038/s41597-022-01710-x},
	doi = {10.1038/s41597-022-01710-x},
	language = {en},
	number = {1},
	urldate = {2024-07-08},
	journal = {Scientific Data},
        author = {Barker,  Michelle and Chue Hong,  Neil P. and Katz,  Daniel S. and Lamprecht,  Anna-Lena and Martinez-Ortiz,  Carlos and Psomopoulos,  Fotis and Harrow,  Jennifer and Castro,  Leyla Jael and Gruenpeter,  Morane and Martinez,  Paula Andrea and Honeyman,  Tom},
	month = oct,
	year = {2022},
	pages = {622},
}

@ARTICLE{huerta2023,
  title     = "{FAIR} for {AI}: An interdisciplinary and international
               community building perspective",
  author = {Huerta,  E. A. and Blaiszik,  Ben and Brinson,  L. Catherine and Bouchard,  Kristofer E. and Diaz,  Daniel and Doglioni,  Caterina and Duarte,  Javier M. and Emani,  Murali and Foster,  Ian and Fox,  Geoffrey and Harris,  Philip and Heinrich,  Lukas and Jha,  Shantenu and Katz,  Daniel S. and Kindratenko,  Volodymyr and Kirkpatrick,  Christine R. and Lassila-Perini,  Kati and Madduri,  Ravi K. and Neubauer,  Mark S. and Psomopoulos,  Fotis E. and Roy,  Avik and R\"{u}bel,  Oliver and Zhao,  Zhizhen and Zhu,  Ruike},
  journal   = "Sci. Data",
  publisher = "Nature Publishing Group",
  volume    =  10,
  number    =  1,
  pages     = "487",
  month     =  jul,
  year      =  2023,
  language  = "en",
  doi = {10.1038/s41597-023-02298-6},
  url = {https://doi.org/10.1038/s41597-023-02298-6}
}

@article{yuan2023,
    author = {Yuan,  David and Ahamed,  Alisha and Burgin,  Josephine and Cummins,  Carla and Devraj,  Rajkumar and Gueye,  Khadim and Gupta,  Dipayan and Gupta,  Vikas and Haseeb,  Muhammad and Ihsan,  Maira and Ivanov,  Eugene and Jayathilaka,  Suran and Kadhirvelu,  Vishnukumar Balavenkataraman and Kumar,  Manish and Lathi,  Ankur and Leinonen,  Rasko and McKinnon,  Jasmine and Meszaros,  Lili and O’Cathail,  Colman and Ouma,  Dennis and Paupério,  Joana and Pesant,  Stephane and Rahman,  Nadim and Rinck,  Gabriele and Selvakumar,  Sandeep and Suman,  Swati and Sunthornyotin,  Yanisa and Ventouratou,  Marianna and Vijayaraja,  Senthilnathan and Waheed,  Zahra and Woollard,  Peter and Zyoud,  Ahmad and Burdett,  Tony and Cochrane,  Guy},
    title = {The European Nucleotide Archive in 2023},
    journal = {Nucleic Acids Research},
    volume = {52},
    number = {D1},
    pages = {D92-D97},
    year = {2023},
    month = {11},
    issn = {0305-1048},
    doi = {10.1093/nar/gkad1067},
    url = {https://doi.org/10.1093/nar/gkad1067},
    eprint = {https://academic.oup.com/nar/article-pdf/52/D1/D92/55040126/gkad1067.pdf},
}

@INPROCEEDINGS{ramapriyan2020,
       author = {{Ramapriyan}, H. and {Behnke}, J.},
        title = "{NASA's Earth Observing System Data and Information System (EOSDIS) and FAIR - A Self-Assessment}",
    booktitle = {AGU Fall Meeting Abstracts},
         year = 2020,
       volume = {2020},
        month = dec,
          eid = {IN044-08},
        pages = {IN044-08},
          url = {https://ui.adsabs.harvard.edu/abs/2020AGUFMIN044..08R}
}

@INPROCEEDINGS{antypas2021,
  author = {Antypas,  K. B. and Bard,  D. J. and Blaschke,  J. P. and Shane Canon,  R. and Enders,  Bjoern and Shankar,  Mallikarjun Arjun and Somnath,  Suhas and Stansberry,  Dale and Uram,  Thomas D. and Wilkinson,  Sean R.},
  booktitle={2021 IEEE International Conference on Big Data (Big Data)}, 
  title={Enabling discovery data science through cross-facility workflows}, 
  year={2021},
  volume={},
  number={},
  pages={3671-3680},
  doi={10.1109/BigData52589.2021.9671421},
  url = {https://doi.org/10.1109/BigData52589.2021.9671421}}

@article{gustafsson2025,
  title = {WorkflowHub: a registry for computational workflows},
  volume = {12},
  ISSN = {2052-4463},
  url = {https://doi.org/10.1038/s41597-025-04786-3},
  DOI = {10.1038/s41597-025-04786-3},
  number = {1},
  journal = {Scientific Data},
  publisher = {Springer Science and Business Media LLC},
  author = {Gustafsson, Ove Johan Ragnar and Wilkinson, Sean R. and Bacall, Finn and Soiland-Reyes, Stian and Leo, Simone and Pireddu, Luca and Owen, Stuart and Juty, Nick and Fern{\'a}ndez, Jos{\'e}M. and Brown, Tom and M{\'e}nager, Herv{\'e} and Gr{\"u}ning, Bj{\"o}rn and Capella-Gutierrez, Salvador and Coppens, Frederik and Goble, Carole},
  year = {2025},
  month = may 
}

@techreport{miller2023,
  author = {Miller,  William and Bard,  Deborah and Boehnlein,  Amber and Fagnan,  Kjiersten and Guok,  Chin and Lan\c{c}on,  Eric and Ramprakash,  Sreeranjani (Jini) and Shankar,  Mallikarjun and Schwarz,  Nicholas and Brown,  Benjamin},
  title        = {Integrated Research Infrastructure Architecture Blueprint Activity (Final Report 2023)},
  institution  = {US Department of Energy (USDOE), Washington, DC (United States). Office of Science; Lawrence Berkeley National Laboratory (LBNL), Berkeley, CA (United States)},
  doi          = {10.2172/1984466},
  url          = {https://www.osti.gov/biblio/1984466},
  place        = {United States},
  year         = {2023},
  month        = {07}}

@misc{fasst2025,
  author       = {{U.S. Department of Energy}},
  title        = {Frontiers in Artificial Intelligence for Science, Security, and Technology Initiative},
  year         = {2025},
  url          = {https://www.energy.gov/science-innovation/energy-science-research/frontiers-artificial-intelligence-science-security-and},
  note         = {Accessed: 2025-03-27}
}

@misc{nairr2025,
  author       = {{U.S. Department of Energy}},
  title        = {National Artificial Intelligence Research Resource},
  year         = {2025},
  url          = {https://www.energy.gov/science-innovation/national-artificial-intelligence-research-resource},
  note         = {Accessed: 2025-03-27}
}

@BOOK{nasem2018,
  author    = {{National Academies of Sciences, Engineering, and Medicine}},
  title     = "Open Science by Design: Realizing a Vision for 21st Century Research",
  isbn      = "978-0-309-47624-9",
  doi       = "10.17226/25116",
  url       = "https://nap.nationalacademies.org/catalog/25116/open-science-by-design-realizing-a-vision-for-21st-century",
  year      = 2018,
  publisher = "The National Academies Press",
  address   = "Washington, DC"
}

@INPROCEEDINGS{stansberry2019,
  author={Stansberry, Dale and Somnath, Suhas and Breet, Jessica and Shutt, Gregory and Shankar, Mallikarjun},
  booktitle={2019 International Conference on Computational Science and Computational Intelligence (CSCI)}, 
  title={DataFed: Towards Reproducible Research via Federated Data Management}, 
  year={2019},
  volume={},
  number={},
  pages={1312-1317},
  doi={10.1109/CSCI49370.2019.00245},
  url = {https://doi.org/10.1109/CSCI49370.2019.00245}}

@INPROCEEDINGS{globus,
  author={Foster, I. and Kesselman, C.},
  booktitle={Proceedings Seventh Heterogeneous Computing Workshop (HCW'98)}, 
  title={The Globus project: a status report}, 
  year={1998},
  volume={},
  number={},
  pages={4-18},
  doi={10.1109/HCW.1998.666541},
  url = {https://doi.org/10.1109/HCW.1998.666541}}

@INPROCEEDINGS{souza2023,
  author={Souza, Renan and Skluzacek, Tyler J. and Wilkinson, Sean R. and Ziatdinov, Maxim and da Silva, Rafael Ferreira},
  booktitle={2023 IEEE 19th International Conference on e-Science (e-Science)}, 
  title={Towards Lightweight Data Integration Using Multi-Workflow Provenance and Data Observability}, 
  year={2023},
  volume={},
  number={},
  pages={1-10},
  doi={10.1109/e-Science58273.2023.10254822},
  url = {https://doi.org/10.1109/e-Science58273.2023.10254822}}

@techreport{wcs2022,
  author = {Ferreira da Silva, Rafael and Badia, Rosa M. and Bala, Venkat and Bard, Debbie and Bremer, Timo and Buckley, Ian and Caino-Lores, Silvina and Chard, Kyle and Goble, Carole and Jha, Shantenu and Katz, Daniel S. and Laney, Daniel and Parashar, Manish and Suter, Frederic and Tyler, Nick and Uram, Thomas and Altintas, Ilkay and Andersson, Stefan and Arndt, William and Aznar, Juan and Bader, Jonathan and Balis, Bartosz and Blanton, Chris and Braghetto, Kelly Rosa and Brodutch, Aharon and Brunk, Paul and Casanova, Henri and Lierta, Alba Cervera and Chigu, Justin and Coleman, Taina and Collier, Nick and Colonnelli, Iacopo and Coppens, Frederik and Crusoe, Michael and Cunningham, Will and De Paula Kinoshita, Bruno and Di Tommaso, Paolo and Doutriaux, Charles and Downton, Matthew and Elwasif, Wael and Enders, Bjoern and Erdmann, Chris and Fahringer, Thomas and Figueiredo, Ludmilla and Filgueira, Rosa and Foltin, Martin and Fouilloux, Anne and Gadelha, Luiz and Gallo, Andy and Saez, Artur Garcia and Garijo, Daniel and Gerlach, Roman and Grant, Ryan and Grayson, Samuel and Grubel, Patricia and Gustafsson, Johan and Hayot-Sasson, Valerie and Hernandez, Oscar and Hilbrich, Marcus and {AnnMary Justine} and Laflotte, Ian and Lehmann, Fabian and Luckow, Andre and Luettgau, Jakob and Maheshwari, Ketan and Matsuda, Motohiko and Medic, Doriana and Mendygral, Pete and Michalewicz, Marek and {Jorji Nonaka} and Pawlik, Maciej and Pottier, Loic and Pouchard, Line and Putz, Mathias and Radha, Santosh Kumar and Ramakrishnan, Lavanya and Ristov, Sashko and Romano, Paul and Rosendo, Daniel and Ruefenacht, Martin and Rycerz, Katarzyna and {Nishant Saurabh} and Savchenko, Volodymyr and Schulz, Martin and Simpson, Christine and Sirvent, Raul and Skluzacek, Tyler and Soiland-Reyes, Stian and Souza, Renan and Sukumar, Sreenivas Rangan and Sun, Ziheng and Sussman, Alan and Thain, Douglas and Titov, Mikhail and Tovar, Benjamin and Tripathy, Aalap and Turilli, Matteo and Tuznik, Bartosz and Van Dam, Hubertus and Vivas, Aurelio and Ward, Logan and Widener, Patrick and Wilkinson, Sean and Zawalska, Justyna and {Mahnoor Zulfiqar}},
  title = {{Workflows Community Summit 2022: A Roadmap Revolution}},
  month = mar,
  year = {2023},
  publisher = {Zenodo},
  number = {ORNL/TM-2023/2885},
  doi = {10.5281/zenodo.7750670},
  url = {https://doi.org/10.5281/zenodo.7750670},
  institution = {Oak Ridge National Laboratory}
}

@techreport{wcs2024,
  author = {Ferreira da Silva,  Rafael and Bard,  Deborah and Chard,  Kyle and Shaun,  de Witt and Foster,  Ian T. and Gibbs,  Tom and Goble,  Carole and Godoy,  William and Gustafsson,  Johan and Haus,  Utz-Uwe and Hudson,  Stephen and Jha,  Shantenu and Los,  Laila and Paine,  Drew and Suter,  Frédéric and Ward,  Logan and Wilkinson,  Sean and Amaris,  Marcos and Babuji,  Yadu and Bader,  Jonathan and Balin,  Riccardo and Balouek,  Daniel and Beecroft,  Sarah and Belhajjame,  Khalid and Bhattarai,  Rajat and Brewer,  Wes and Brunk,  Paul and Caino-Lores,  Silvina and Casanova,  Henri and Cassol,  Daniela and Coleman,  Jared and Coleman,  Taina and Colonnelli,  Iacopo and Da Silva,  Anderson Andrei and de Oliveira,  Daniel and Elahi,  Pascal and Elfaramawy,  Nour and Elwasif,  Wael and Etz,  Brian and Fahringer,  Thomas and Ferreira,  Wesley and Filgueira,  Rosa and Fosso Tande,  Jacob and Gadelha,  Luiz and Gallo,  Andy and Garijo,  Daniel and Georgiou,  Yiannis and Gritsch,  Philipp and Grubel,  Patricia and Gueroudji,  Amal and Guilloteau,  Quentin and Hamalainen,  Carlo and Hong Enriquez,  Rolando and Huet,  Lauren and Hunter Kesling,  Kevin and Iborra,  Paula and Jahangiri,  Shiva and Janssen,  Jan and Jordan,  Joe and Kanwal,  Sehrish and Kunstmann,  Liliane and Lehmann,  Fabian and Leser,  Ulf and Li,  Chen and Liu,  Peini and Luettgau,  Jakob and Lupat,  Richard and M. Fernandez,  Jose and Maheshwari,  Ketan and Malik,  Tanu and Marquez,  Jack and Matsuda,  Motohiko and Medic,  Doriana and Mohammadi,  Somayeh and Mulone,  Alberto and Navarro,  John-Luke and Ng,  Kin Wai and Noelp,  Klaus and P. Kinoshita,  Bruno and Prout,  Ryan and R. Crusoe,  Michael and Ristov,  Sashko and Robila,  Stefan and Rosendo,  Daniel and Rowell,  Billy and Rybicki,  Jedrzej and Sanchez,  Hector and Saurabh,  Nishant and Saurav,  Sumit Kumar and Scogland,  Tom and Senanayake,  Dinindu and Shin,  Woong and Sirvent,  Raul and Skluzacek,  Tyler and Sly-Delgado,  Barry and Soiland-Reyes,  Stian and Souza,  Abel and Souza,  Renan and Talia,  Domenico and Tallent,  Nathan and Thamsen,  Lauritz and Titov,  Mikhail and Tovar,  Benjamin and Vahi,  Karan and Vardar-Irrgang,  Eric and Vartina,  Edite and Wang,  Yuandou and Wouters,  Merridee and Yu,  Qi and Al Bkhetan,  Ziad and Zulfiqar,  Mahnoor},
title = {{Workflows Community Summit 2024: Future Trends and Challenges in Scientific Workflows}}, year = {2024},
publisher = {Zenodo},
number = {ORNL/TM-2024/3573},
doi = {10.5281/zenodo.13844759},
url = {https://doi.org/10.5281/zenodo.13844759},
institution = {Oak Ridge National Laboratory} }

@article{soiland-reyes2022,
	title = {Packaging research artefacts with {RO}-{Crate}},
	volume = {5},
	copyright = {https://creativecommons.org/licenses/by/4.0/},
	issn = {24518492, 24518484},
	url = {https://doi.org/10.3233/DS-210053},
	doi = {10.3233/DS-210053},
	number = {2},
	urldate = {2024-08-07},
	journal = {Data Science},
	author = {Soiland-Reyes, Stian and Sefton, Peter and Crosas, Mercè and Castro, Leyla Jael and Coppens, Frederik and Fernández, José M. and others},
	editor = {Peroni, Silvio},
	month = jul,
	year = {2022},
	pages = {97--138},
}

@article{leo2024,
  title={Recording provenance of workflow runs with RO-Crate},
  author = {Leo,  Simone and Crusoe,  Michael R. and Rodríguez-Navas,  Laura and Sirvent,  Ra\"{u}l and Kanitz,  Alexander and De Geest,  Paul and Wittner,  Rudolf and Pireddu,  Luca and Garijo,  Daniel and Fernández,  José M. and Colonnelli,  Iacopo and Gallo,  Matej and Ohta,  Tazro and Suetake,  Hirotaka and Capella-Gutierrez,  Salvador and de Wit,  Renske and Kinoshita,  Bruno P. and Soiland-Reyes,  Stian},
  journal={PLoS one},
  volume={19},
  number={9},
  pages={e0309210},
  year={2024},
  publisher={Public Library of Science San Francisco, CA USA},
  url = {https://doi.org/10.1371/journal.pone.0309210}
}

@inproceedings{gray2017,
title = "Bioschemas: From Potato Salad to Protein Annotation",
author = "Gray, {Alasdair J G} and Carole Goble and Jimenez, {Rafael C.}",
year = "2017",
month = oct,
day = "22",
language = "English",
series = "CEUR workshop proceedings",
publisher = "RWTH Aachen University",
editor = "Nadeschda Nikitina and Dezhao Song and Fokoue, {Achille } and Haase, { Peter}",
booktitle = "ISWC 2017 Posters \& Demonstrations and Industry Tracks",
address = "Germany",
edition = "urn:nbn:de:0074-1963-7",
note = "The 16th International Semantic Web Conference 2017, ISWC 2017 ; Conference date: 21-10-2018 Through 25-10-2018",
url = "https://iswc2017.semanticweb.org/paper-579/",
}

@article{crusoe2022,
	title = {Methods included: standardizing computational reuse and portability with the {Common} {Workflow} {Language}},
	volume = {65},
	issn = {0001-0782, 1557-7317},
	shorttitle = {Methods included},
	url = {https://doi.org/10.1145/3486897},
	doi = {10.1145/3486897},
	language = {en},
	number = {6},
	urldate = {2024-08-07},
	journal = {Communications of the ACM},
        author = {Crusoe,  Michael R. and Abeln,  Sanne and Iosup,  Alexandru and Amstutz,  Peter and Chilton,  John and Tijanić,  Nebojša and Ménager,  Hervé and Soiland-Reyes,  Stian and Gavrilović,  Bogdan and Goble,  Carole and The CWL Community},
	month = jun,
	year = {2022},
	pages = {54--63},
}

@misc{miljkovic2022,
  author       = {Miljković, Nadica and
                  Trisovic, Ana and
                  Peer, Limor},
  title        = {Towards {FAIR Principles} for Open Hardware},
  month        = feb,
  year         = 2022,
  publisher    = {University of Belgrade - School of Electrical
                   Engineering and Academic Mind
                  },
  version      = 3,
  doi          = {10.5281/zenodo.6506428},
  url          = {https://doi.org/10.5281/zenodo.6506428},
}

@techreport{johnson2024,
  doi = {10.5065/ZGSX-2D06},
  url = {https://opensky.ucar.edu/islandora/object/technotes:601},
  author = {Johnson,  Andrew and Julian,  Renaine and Mayernik,  Matthew and Mundoma,  Claudius and Murray,  Matthew and Ranganath,  Aditya and Stossmeister,  Gregory},
  title = {{FAIR Facilities and Instruments Workshop \#1 Report: Exploring Persistent Identifier Needs,  Barriers,  and Incentives}},
  publisher = {National Center for Atmospheric Research},
  institution = {National Center for Atmospheric Research},
  year = {2024}
}

@conference{wolf2021,
  author       = {Wolf, Matthew and Logan, Jeremy and Mehta, Kshitij and Jacobson, Daniel A. and McDevitt, Mikaela Cashman and Walker, Angelica and Eisenhauer, Greg and Widener, Patrick and Cliff, Ashley},
  title        = {{Reusability First: Toward FAIR Workflows}},
  url          = {https://www.osti.gov/biblio/1827005},
  place        = {United States},
  organization = {Oak Ridge National Lab. (ORNL), Oak Ridge, TN (United States)},
  year         = {2021},
  month        = {09}}

@techreport{caw2021,
  author       = {Wilkinson, Sean and Knight, Kathryn and Kuchar, Olga and Mehta, Kshitij and Shankar, Mallikarjun and Wolf, Matthew},
  title        = {Official Report on the 2021 Computational and Autonomous Workflows Workshop ({CAW} 2021)},
  institution  = {Oak Ridge National Laboratory (ORNL), Oak Ridge, TN (United States)},
  doi          = {10.2172/1862119},
  url          = {https://www.osti.gov/biblio/1862119},
  place        = {United States},
  year         = {2022},
  month        = {03}}

@article{hasselbring2019,
  author       = {Wilhelm Hasselbring and
                  Leslie Carr and
                  Simon Hettrick and
                  Heather S. Packer and
                  Thanassis Tiropanis},
  title        = {{FAIR and Open Computer Science Research Software}},
  journal      = {CoRR},
  volume       = {abs/1908.05986},
  year         = {2019},
  url          = {http://arxiv.org/abs/1908.05986},
  eprinttype    = {arXiv},
  eprint       = {1908.05986},
  bibsource    = {dblp computer science bibliography, https://dblp.org}
}

@article{carroll2020,
  title = {The CARE Principles for Indigenous Data Governance},
  volume = {19},
  ISSN = {1683-1470},
  url = {http://doi.org/10.5334/dsj-2020-043},
  DOI = {10.5334/dsj-2020-043},
  journal = {Data Science Journal},
  publisher = {Ubiquity Press,  Ltd.},
  author = {Carroll,  Stephanie Russo and Garba,  Ibrahim and Figueroa-Rodríguez,  Oscar L. and Holbrook,  Jarita and Lovett,  Raymond and Materechera,  Simeon and Parsons,  Mark and Raseroka,  Kay and Rodriguez-Lonebear,  Desi and Rowe,  Robyn and Sara,  Rodrigo and Walker,  Jennifer D. and Anderson,  Jane and Hudson,  Maui},
  year = {2020}
}

@article{jordan2024,
  title = {SHARE: A Framework for Secondary Qualitative Data Analysis},
  volume = {5},
  ISSN = {2690-5450},
  url = {http://doi.org/10.21061/see.175},
  DOI = {10.21061/see.175},
  number = {1},
  journal = {Studies in Engineering Education},
  publisher = {Virginia Tech Libraries},
  author = {Jordan,  Shawn S. and Matusovich,  Holly M. and Case,  Jennifer M. and Benson,  Lisa and Delaine,  David A. and Kajfez,  Rachel L. and Lord,  Susan M. and Paretti,  Marie C. and Young,  E. Tyler and Zastavker,  Yevgeniya V.},
  year = {2024},
  pages = {125–133}
}

@techreport{share2018,
  author        = {},
  title         = {Principles for Conducting Research in the Arctic},
  number =        {},
  institution =   {Interagency Arctic Research Policy Committee},
  address =       {Washington D.C},
  year          = {2018},
  url = {https://www.iarpccollaborations.org/uploads/cms/documents/principles_for_conducting_research_in_the_arctic_final_2018.pdf}
}

@article{lin2020,
  title = {The {TRUST Principles} for digital repositories},
  volume = {7},
  ISSN = {2052-4463},
  url = {http://doi.org/10.1038/s41597-020-0486-7},
  DOI = {10.1038/s41597-020-0486-7},
  number = {1},
  journal = {Scientific Data},
  publisher = {Springer Science and Business Media LLC},
  author = {Lin,  Dawei and Crabtree,  Jonathan and Dillo,  Ingrid and Downs,  Robert R. and Edmunds,  Rorie and Giaretta,  David and De Giusti,  Marisa and L’Hours,  Hervé and Hugo,  Wim and Jenkyns,  Reyna and Khodiyar,  Varsha and Martone,  Maryann E. and Mokrane,  Mustapha and Navale,  Vivek and Petters,  Jonathan and Sierman,  Barbara and Sokolova,  Dina V. and Stockhause,  Martina and Westbrook,  John},
  year = {2020},
  month = may 
}

@misc{wilkinson2025-2,
      title={{An Ecosystem of Services for FAIR Computational Workflows}}, 
      author={Sean R. Wilkinson and Johan Gustafsson and Finn Bacall and Khalid Belhajjame and Salvador Capella and Jose Maria Fernandez Gonzalez and Jacob Fosso Tande and Luiz Gadelha and Daniel Garijo and Patricia Grubel and Bjorn Grüning and Farah Zaib Khan and Sehrish Kanwal and Simone Leo and Stuart Owen and Luca Pireddu and Line Pouchard and Laura Rodríguez-Navas and Beatriz Serrano-Solano and Stian Soiland-Reyes and Baiba Vilne and Alan Williams and Merridee Ann Wouters and Frederik Coppens and Carole Goble},
      year={2025},
      eprint={2505.15988},
      archivePrefix={arXiv},
      primaryClass={cs.DC},
      doi = {10.48550/arXiv.2505.15988},
      url={https://arxiv.org/abs/2505.15988}, 
}

@inproceedings{wilkinson2025-3,
author = {Wilkinson, Sean R. and Widener, Patrick},
title = {{FAIR Ecosystems for Science at Scale}},
year = {2025},
isbn = {9798400713989},
publisher = {Association for Computing Machinery},
address = {New York, NY, USA},
url = {https://doi.org/10.1145/3708035.3736097},
doi = {10.1145/3708035.3736097},
booktitle = {Practice and Experience in Advanced Research Computing 2025: The Power of Collaboration},
articleno = {47},
numpages = {4},
location = {},
series = {PEARC '25}
}

@INPROCEEDINGS{vazhkudai2016,
  author={Vazhkudai, Sudharshan S. and Harney, John and Gunasekaran, Raghul and Stansberry, Dale and Lim, Seung-Hwan and Barron, Tom and Nash, Andrew and Ramanathan, Arvind},
  booktitle={2016 IEEE International Conference on Big Data (Big Data)}, 
  title={Constellation: A science graph network for scalable data and knowledge discovery in extreme-scale scientific collaborations}, 
  year={2016},
  volume={},
  number={},
  pages={3052-3061},
  doi={10.1109/BigData.2016.7840959},
    url = {https://doi.org/10.1109/BigData.2016.7840959}
}

@article{suter2025,
title = {A terminology for scientific workflow systems},
journal = {Future Generation Computer Systems},
volume = {174},
pages = {107974},
year = {2026},
issn = {0167-739X},
doi = {https://doi.org/10.1016/j.future.2025.107974},
url = {https://www.sciencedirect.com/science/article/pii/S0167739X25002699},
author = {Frédéric Suter and Tainã Coleman and İlkay Altintaş and Rosa M. Badia and Bartosz Balis and Kyle Chard and Iacopo Colonnelli and Ewa Deelman and Paolo {Di Tommaso} and Thomas Fahringer and Carole Goble and Shantenu Jha and Daniel S. Katz and Johannes Köster and Ulf Leser and Kshitij Mehta and Hilary Oliver and J.-Luc Peterson and Giovanni Pizzi and Loïc Pottier and Raül Sirvent and Eric Suchyta and Douglas Thain and Sean R. Wilkinson and Justin M. Wozniak and Rafael {Ferreira da Silva}},
}

@article{ison2015,
    author = {Ison, Jon and Rapacki, Kristoffer and Ménager, Hervé and Kalaš, Matúš and Rydza, Emil and Chmura, Piotr and Anthon, Christian and Beard, Niall and Berka, Karel and Bolser, Dan and Booth, Tim and Bretaudeau, Anthony and Brezovsky, Jan and Casadio, Rita and Cesareni, Gianni and Coppens, Frederik and Cornell, Michael and Cuccuru, Gianmauro and Davidsen, Kristian and Vedova, Gianluca Della and Dogan, Tunca and Doppelt-Azeroual, Olivia and Emery, Laura and Gasteiger, Elisabeth and Gatter, Thomas and Goldberg, Tatyana and Grosjean, Marie and Grüning, Björn and Helmer-Citterich, Manuela and Ienasescu, Hans and Ioannidis, Vassilios and Jespersen, Martin Closter and Jimenez, Rafael and Juty, Nick and Juvan, Peter and Koch, Maximilian and Laibe, Camille and Li, Jing-Woei and Licata, Luana and Mareuil, Fabien and Mičetić, Ivan and Friborg, Rune Møllegaard and Moretti, Sebastien and Morris, Chris and Möller, Steffen and Nenadic, Aleksandra and Peterson, Hedi and Profiti, Giuseppe and Rice, Peter and Romano, Paolo and Roncaglia, Paola and Saidi, Rabie and Schafferhans, Andrea and Schwämmle, Veit and Smith, Callum and Sperotto, Maria Maddalena and Stockinger, Heinz and Vařeková, Radka Svobodová and Tosatto, Silvio C.E. and de la Torre, Victor and Uva, Paolo and Via, Allegra and Yachdav, Guy and Zambelli, Federico and Vriend, Gert and Rost, Burkhard and Parkinson, Helen and Løngreen, Peter and Brunak, Søren},
    title = {Tools and data services registry: a community effort to document bioinformatics resources},
    journal = {Nucleic Acids Research},
    volume = {44},
    number = {D1},
    pages = {D38-D47},
    year = {2015},
    month = {11},
    issn = {0305-1048},
    doi = {10.1093/nar/gkv1116},
    url = {https://doi.org/10.1093/nar/gkv1116},
    eprint = {https://academic.oup.com/nar/article-pdf/44/D1/D38/9482499/gkv1116.pdf},
}

@article{ison2019,
  title = {The bio.tools registry of software tools and data resources for the life sciences},
  volume = {20},
  ISSN = {1474-760X},
  url = {https://doi.org/10.1186/s13059-019-1772-6},
  DOI = {10.1186/s13059-019-1772-6},
  number = {1},
  journal = {Genome Biology},
  publisher = {Springer Science and Business Media LLC},
  author = {Ison,  Jon and Ienasescu,  Hans and Chmura,  Piotr and Rydza,  Emil and Ménager,  Hervé and Kalaš,  Matúš and Schw\"{a}mmle,  Veit and Gr\"{u}ning,  Bj\"{o}rn and Beard,  Niall and Lopez,  Rodrigo and Duvaud,  Severine and Stockinger,  Heinz and Persson,  Bengt and Vařeková,  Radka Svobodová and Raček,  Tomáš and Vondrášek,  Jiří and Peterson,  Hedi and Salumets,  Ahto and Jonassen,  Inge and Hooft,  Rob and Nyr\"{o}nen,  Tommi and Valencia,  Alfonso and Capella,  Salvador and Gelpí,  Josep and Zambelli,  Federico and Savakis,  Babis and Leskošek,  Brane and Rapacki,  Kristoffer and Blanchet,  Christophe and Jimenez,  Rafael and Oliveira,  Arlindo and Vriend,  Gert and Collin,  Olivier and van Helden,  Jacques and Løngreen,  Peter and Brunak,  Søren},
  year = {2019},
  month = aug 
}

@article{daveigaleprevost2017,
  title = {BioContainers: an open-source and community-driven framework for software standardization},
  volume = {33},
  ISSN = {1367-4811},
  url = {https://doi.org/10.1093/bioinformatics/btx192},
  DOI = {10.1093/bioinformatics/btx192},
  number = {16},
  journal = {Bioinformatics},
  publisher = {Oxford University Press (OUP)},
  author = {da Veiga Leprevost,  Felipe and Gr\"{u}ning,  Bj\"{o}rn A and Alves Aflitos,  Saulo and R\"{o}st,  Hannes L and Uszkoreit,  Julian and Barsnes,  Harald and Vaudel,  Marc and Moreno,  Pablo and Gatto,  Laurent and Weber,  Jonas and Bai,  Mingze and Jimenez,  Rafael C and Sachsenberg,  Timo and Pfeuffer,  Julianus and Vera Alvarez,  Roberto and Griss,  Johannes and Nesvizhskii,  Alexey I and Perez-Riverol,  Yasset},
  editor = {Valencia,  Alfonso},
  year = {2017},
  month = mar,
  pages = {2580–2582}
}

@techreport{julian2024,
  doi = {10.5065/JEA7-YF24},
  url = {https://opensky.ucar.edu/islandora/object/technotes:42004},
  author = {Julian,  Renaine and Johnson,  Andrew and Mayernik,  Matthew and Mundoma,  Claudius and Murray,  Matthew and Ranganath,  Aditya},
  title = {{FAIR Facilities and Instruments Workshop \#2 Report: Recent Progress,  Remaining Challenges,  and Emerging PID Strategies}},
  institution = {NSF National Center for Atmospheric Research},
  year = {2024}
}

@inproceedings{skluzacek2025,
author = {Skluzacek, Tyler J. and Bryant, Paul and Ruckman, A.J. and Rosendo, Daniel and Prentice, Suzanne and Brim, Michael J. and Adamson, Ryan and Oral, Sarp and Shankar, Mallikarjun and Ferreira da Silva, Rafael},
title = {Secure API-Driven Research Automation to Accelerate Scientific Discovery},
year = {2025},
isbn = {9798400713989},
publisher = {Association for Computing Machinery},
address = {New York, NY, USA},
url = {https://doi.org/10.1145/3708035.3736072},
doi = {10.1145/3708035.3736072},
booktitle = {Practice and Experience in Advanced Research Computing 2025: The Power of Collaboration},
articleno = {68},
numpages = {5},
location = {
},
series = {PEARC '25}
}

@article{rehm2021,
  title = {GA4GH: International policies and standards for data sharing across genomic research and healthcare},
  volume = {1},
  ISSN = {2666-979X},
  url = {https://doi.org/10.1016/j.xgen.2021.100029},
  DOI = {10.1016/j.xgen.2021.100029},
  number = {2},
  journal = {Cell Genomics},
  publisher = {Elsevier BV},
  author = {Rehm,  Heidi L. and Page,  Angela J.H. and Smith,  Lindsay and Adams,  Jeremy B. and Alterovitz,  Gil and Babb,  Lawrence J. and Barkley,  Maxmillian P. and Baudis,  Michael and Beauvais,  Michael J.S. and Beck,  Tim and Beckmann,  Jacques S. and Beltran,  Sergi and Bernick,  David and Bernier,  Alexander and Bonfield,  James K. and Boughtwood,  Tiffany F. and Bourque,  Guillaume and Bowers,  Sarion R. and Brookes,  Anthony J. and Brudno,  Michael and Brush,  Matthew H. and Bujold,  David and Burdett,  Tony and Buske,  Orion J. and Cabili,  Moran N. and Cameron,  Daniel L. and Carroll,  Robert J. and Casas-Silva,  Esmeralda and Chakravarty,  Debyani and Chaudhari,  Bimal P. and Chen,  Shu Hui and Cherry,  J. Michael and Chung,  Justina and Cline,  Melissa and Clissold,  Hayley L. and Cook-Deegan,  Robert M. and Courtot,  Mélanie and Cunningham,  Fiona and Cupak,  Miro and Davies,  Robert M. and Denisko,  Danielle and Doerr,  Megan J. and Dolman,  Lena I. and Dove,  Edward S. and Dursi,  L. Jonathan and Dyke,  Stephanie O.M. and Eddy,  James A. and Eilbeck,  Karen and Ellrott,  Kyle P. and Fairley,  Susan and Fakhro,  Khalid A. and Firth,  Helen V. and Fitzsimons,  Michael S. and Fiume,  Marc and Flicek,  Paul and Fore,  Ian M. and Freeberg,  Mallory A. and Freimuth,  Robert R. and Fromont,  Lauren A. and Fuerth,  Jonathan and Gaff,  Clara L. and Gan,  Weiniu and Ghanaim,  Elena M. and Glazer,  David and Green,  Robert C. and Griffith,  Malachi and Griffith,  Obi L. and Grossman,  Robert L. and Groza,  Tudor and Guidry Auvil,  Jaime M. and Guigó,  Roderic and Gupta,  Dipayan and Haendel,  Melissa A. and Hamosh,  Ada and Hansen,  David P. and Hart,  Reece K. and Hartley,  Dean Mitchell and Haussler,  David and Hendricks-Sturrup,  Rachele M. and Ho,  Calvin W.L. and Hobb,  Ashley E. and Hoffman,  Michael M. and Hofmann,  Oliver M. and Holub,  Petr and Hsu,  Jacob Shujui and Hubaux,  Jean-Pierre and Hunt,  Sarah E. and Husami,  Ammar and Jacobsen,  Julius O. and Jamuar,  Saumya S. and Janes,  Elizabeth L. and Jeanson,  Francis and Jené,  Aina and Johns,  Amber L. and Joly,  Yann and Jones,  Steven J.M. and Kanitz,  Alexander and Kato,  Kazuto and Keane,  Thomas M. and Kekesi-Lafrance,  Kristina and Kelleher,  Jerome and Kerry,  Giselle and Khor,  Seik-Soon and Knoppers,  Bartha M. and Konopko,  Melissa A. and Kosaki,  Kenjiro and Kuba,  Martin and Lawson,  Jonathan and Leinonen,  Rasko and Li,  Stephanie and Lin,  Michael F. and Linden,  Mikael and Liu,  Xianglin and Liyanage,  Isuru Udara and Lopez,  Javier and Lucassen,  Anneke M. and Lukowski,  Michael and Mann,  Alice L. and Marshall,  John and Mattioni,  Michele and Metke-Jimenez,  Alejandro and Middleton,  Anna and Milne,  Richard J. and Molnár-Gábor,  Fruzsina and Mulder,  Nicola and Munoz-Torres,  Monica C. and Nag,  Rishi and Nakagawa,  Hidewaki and Nasir,  Jamal and Navarro,  Arcadi and Nelson,  Tristan H. and Niewielska,  Ania and Nisselle,  Amy and Niu,  Jeffrey and Nyr\"{o}nen,  Tommi H. and O’Connor,  Brian D. and Oesterle,  Sabine and Ogishima,  Soichi and Ota Wang,  Vivian and Paglione,  Laura A.D. and Palumbo,  Emilio and Parkinson,  Helen E. and Philippakis,  Anthony A. and Pizarro,  Angel D. and Prlic,  Andreas and Rambla,  Jordi and Rendon,  Augusto and Rider,  Renee A. and Robinson,  Peter N. and Rodarmer,  Kurt W. and Rodriguez,  Laura Lyman and Rubin,  Alan F. and Rueda,  Manuel and Rushton,  Gregory A. and Ryan,  Rosalyn S. and Saunders,  Gary I. and Schuilenburg,  Helen and Schwede,  Torsten and Scollen,  Serena and Senf,  Alexander and Sheffield,  Nathan C. and Skantharajah,  Neerjah and Smith,  Albert V. and Sofia,  Heidi J. and Spalding,  Dylan and Spurdle,  Amanda B. and Stark,  Zornitza and Stein,  Lincoln D. and Suematsu,  Makoto and Tan,  Patrick and Tedds,  Jonathan A. and Thomson,  Alastair A. and Thorogood,  Adrian and Tickle,  Timothy L. and Tokunaga,  Katsushi and T\"{o}rnroos,  Juha and Torrents,  David and Upchurch,  Sean and Valencia,  Alfonso and Guimera,  Roman Valls and Vamathevan,  Jessica and Varma,  Susheel and Vears,  Danya F. and Viner,  Coby and Voisin,  Craig and Wagner,  Alex H. and Wallace,  Susan E. and Walsh,  Brian P. and Williams,  Marc S. and Winkler,  Eva C. and Wold,  Barbara J. and Wood,  Grant M. and Woolley,  J. Patrick and Yamasaki,  Chisato and Yates,  Andrew D. and Yung,  Christina K. and Zass,  Lyndon J. and Zaytseva,  Ksenia and Zhang,  Junjun and Goodhand,  Peter and North,  Kathryn and Birney,  Ewan},
  year = {2021},
  month = nov,
  pages = {100029}
}

@inproceedings{atchley2023,
  series = {SC ’23},
  title = {Frontier: Exploring Exascale},
  url = {https://doi.org/10.1145/3581784.3607089},
  DOI = {10.1145/3581784.3607089},
  booktitle = {Proceedings of the International Conference for High Performance Computing,  Networking,  Storage and Analysis},
  publisher = {ACM},
  author = {Atchley,  Scott and Zimmer,  Christopher and Lange,  John and Bernholdt,  David and Melesse Vergara,  Veronica and Beck,  Thomas and Brim,  Michael and Budiardja,  Reuben and Chandrasekaran,  Sunita and Eisenbach,  Markus and Evans,  Thomas and Ezell,  Matthew and Frontiere,  Nicholas and Georgiadou,  Antigoni and Glenski,  Joe and Grete,  Philipp and Hamilton,  Steven and Holmen,  John and Huebl,  Axel and Jacobson,  Daniel and Joubert,  Wayne and Mcmahon,  Kim and Merzari,  Elia and Moore,  Stan and Myers,  Andrew and Nichols,  Stephen and Oral,  Sarp and Papatheodore,  Thomas and Perez,  Danny and Rogers,  David M. and Schneider,  Evan and Vay,  Jean-Luc and Yeung,  P. K.},
  year = {2023},
  month = nov,
  pages = {1–16},
  collection = {SC '23}
}

@misc{big-bill,
  title={{One Big Beautiful Bill Act}},
  author={Arrington, Jodey C.},
  howpublished={House Bill H.R. 1, 119th Congress},
  year={2025},
  url={https://www.congress.gov/bill/119th-congress/house-bill/1}
}

@misc{bernerslee2006,
  author       = {Tim Berners-Lee},
  title        = {Linked Data},
  year         = {2006},
  url          = {https://www.w3.org/DesignIssues/LinkedData.html},
  note         = {Accessed: 2025-08-27}
}

@techreport{liao2024,
  author       = {Liao, Chunhua and Shen, Xipeng and Emani, Murali and Vanderbruggen, Tristan and Lin, Pei-Hung},
  title        = {{HPC-FAIR: A Framework Managing Data and AI Models
 for Analyzing and Optimizing Scientific Applications}},
  institution  = {North Carolina State University (NCSU)},
  doi          = {10.2172/2504172},
  url          = {https://www.osti.gov/biblio/2504172},
  place        = {United States},
  year         = {2024},
  month        = {09}}

@inproceedings{valero-lara2024,
title = "{ChatBLAS: The First AI-Generated and Portable BLAS Library}",
author = "Pedro Valero-Lara and Godoy, {William F.} and Keita Teranishi and Prasanna Balaprakash and Vetter, {Jeffrey S.}",
year = "2024",
doi = "10.1109/SCW63240.2024.00010",
url = {https://doi.org/10.1109/SCW63240.2024.00010},
language = "English",
series = "Proceedings of SC 2024-W: Workshops of the International Conference for High Performance Computing, Networking, Storage and Analysis",
publisher = "Institute of Electrical and Electronics Engineers Inc.",
pages = "19--24",
booktitle = "Proceedings of SC 2024-W",
}

@misc{vonlaszewski2025,
      title={Towards Experiment Execution in Support of Community Benchmark Workflows for {HPC}}, 
      author={Gregor von Laszewski and Wesley Brewer and Sean R. Wilkinson and Andrew Shao and J. P. Fleischer and Harshad Pitkar and Christine R. Kirkpatrick and Geoffrey C. Fox},
      year={2025},
      eprint={2507.22294},
      archivePrefix={arXiv},
      primaryClass={cs.DC},
      doi = {10.48550/arXiv.2507.22294},
      url={https://arxiv.org/abs/2507.22294}, 
}

\end{document}